\documentclass[a4paper,11pt]{article}
\usepackage{amsfonts,graphicx}

\input{tcilatex}

\begin{document}

%
%

\setcounter{page}{0} \topmargin0pt \oddsidemargin5mm \renewcommand{%
\thefootnote}{\fnsymbol{footnote}} \newpage \setcounter{page}{0} 
\begin{titlepage}
\begin{flushright}
Berlin Sfb288 Preprint  \\
hep-th/0205076 
\end{flushright}
\vspace{0.2cm}
\begin{center}
{\Large {\bf From integrability  to conductance, impurity systems} }

\vspace{0.8cm}
{\Large  Olalla~Castro-Alvaredo  and  Andreas~Fring }

\vspace{0.2cm}
{Institut f\"ur Theoretische Physik, 
Freie Universit\"at Berlin,\\
Arnimallee 14, D-14195 Berlin, Germany }
\end{center}
\vspace{0.5cm}
 
\renewcommand{\thefootnote}{\arabic{footnote}}
\setcounter{footnote}{0}

\begin{abstract}
We compute the DC conductance with two different methods, which both
exploit the integrability of the theories under consideration. On one hand we
determine the conductance through a defect by means of the thermodynamic 
Bethe ansatz and standard relativistic potential scattering theory based on a Landauer 
transport theory picture. On the other hand we propose a Kubo formula for a defect 
system  and evaluate the current-current two-point 
correlation function it involves with the help of a form factor expansion. For a variety of 
defects in a fermionic system we find excellent agreement between the two 
different theoretical descriptions.

\par\noindent
PACS numbers: 73.50.Bk, 73.40.-c, 72.10.-d, 11.10Kk, 05.30.-d
\end{abstract}
\vfill{ \hspace*{-9mm}
\begin{tabular}{l}
\rule{6 cm}{0.05 mm}\\
Olalla@physik.fu-berlin.de\\
Fring@physik.fu-berlin.de
\end{tabular}}
\end{titlepage}
\newpage 
\tableofcontents
\newpage 

\section{Introduction}

\vspace{-0.2cm} Conductance (conductivity) measurements belong to the
easiest and most direct experiments which can be carried out. They attract a
lot of attention, due to the fact that in general they can be performed
without perturbing very much the behaviour of the system, e.g. a
rigid-lattice bulk metal, such that the uncertainty of experimental
artefacts is reduced to a minimum. There exist various well-known
theoretical descriptions, such as semi-classical transport theories
(Landauer \cite{Land} and Boltzmann-Drude \cite{BD}), dynamical
linear-response theory \cite{kubo,KTH} and also Green function
linear-response theory \cite{Fen}. To carry out the latter, in particular at
finite temperature, is still poorly understood in generality \cite{Jeon},
even in 1+1 space-time dimensions \cite{Ray}. Since recent experimental
progress allows conductance measurements also in 1+1 space-time dimensions 
\cite{Mill}, one can on the theoretical side fully exploit the special
features of low dimensionality.

It is in particular very suggestive to exploit the full scope of
non-perturbative techniques which have been developed in the context of
integrable quantum field theories in 1+1 space-time dimensions, such as the
thermodynamic Bethe ansatz (TBA) \cite{TBAZam,TBAKM} and the form factor
bootstrap approach \cite{KW,Smir}. Generalizing the Landauer transport
picture a proposal for the conductance through a quantum wire with a defect
(impurity) has been made in \cite{FLS,FLS2} 
\begin{equation}
G^{\mathbf{\alpha }}(T)=\sum_{i}\lim_{(\mu _{i}^{l}-\mu _{i}^{r})\rightarrow
0}\frac{q_{i}}{2}\int\limits_{-\infty }^{\infty }d\theta \left[ \rho
_{i}^{r}(\theta ,T,\mu _{i}^{l})|T_{i}^{\mathbf{\alpha }}\left( \theta
\right) |^{2}-\rho _{i}^{r}(\theta ,T,\mu _{i}^{r})|\tilde{T}_{i}^{\mathbf{%
\alpha }}\left( \theta \right) |^{2}\right] ,  \label{1}
\end{equation}
which we only modify to accommodate parity breaking, known to occur in
integrable lattice models, see e.g. \cite{Kadar}. This means in particular
we allow the transmission amplitudes to be different for a particle of type $%
i$ with charge $q_{i}$ passing with rapidity $\theta $ through a defect of
type $\mathbf{\alpha }$ from the left $T_{i}^{\mathbf{\alpha }}\left( \theta
\right) $ and right $\tilde{T}_{i}^{\mathbf{\alpha }}\left( \theta \right) $%
. The density distribution function $\rho _{i}^{r}(\theta ,T,\mu _{i})$,
being a function the temperature $T$, and the potential at the left $\mu
_{i}^{l}$ and right $\mu _{i}^{r}$ constriction of the wire, can be
determined by means of the TBA. We have already restricted (\ref{1}) to the
abelian (diagonal) situation. It is clear that the effect resulting from the
defect is most interesting when $|T_{i}^{\mathbf{\alpha }}\left( \theta
\right) |\neq 1$, which requires the occurrence of simultaneous transmission
and reflection (see (\ref{U2}), (\ref{c2})). In this paper we will therefore
be mainly interested in that situation. One may adapt (\ref{1}) also to the
case of pure reflection, which physically describes the influence of the
constriction to the conducting process. From the previous statement it is
clear that such boundary theories are only interesting in this physical
context when they are non-abelian.

The other prominent way of determining the conductance is a result from
linear response theory, which yields an expression for the conductance in
form of the Fourier transform of the current-current two-point correlation
function. This Kubo formula has been adopted to the situation with a
boundary \cite{LSS}. As we mentioned, this will only capture effects coming
from the constriction of the wire, we propose here a generalization to the
analogous situation as described in (\ref{1}), i.e. when a defect is present

\begin{equation}
G^{\mathbf{\alpha }}(T)=-\lim_{\omega \rightarrow 0}\frac{1}{2\omega \pi ^{2}%
}\int\limits_{-\infty }^{\infty }dt\,\,e^{i\omega t}\,\,\left\langle J(t)Z_{%
\mathbf{\alpha }}\,J(0)\right\rangle _{T,m}.  \label{2}
\end{equation}
Here the defect operator $Z_{\mathbf{\alpha }}$ enters in-between the two
currents $J$ within the temperature and mass $m$ dependent correlation
function. The Matsubara frequency is denoted by $\omega $.

The main purpose of this manuscript is to compare the two alternative
descriptions (\ref{1}) and (\ref{2}) for massive bulk theories with a defect
which allows for simultaneous reflection and transmission. There exist
various investigations, e.g., \cite{WA,FLS,FLS2,pureT} for conformal
(massless) theories with defect, which exploit the original folding idea of
Wong and Affleck \cite{WA}. The idea is that a conformal field theory with a 
\emph{purely} transmitting or reflecting defect can be mapped into a
boundary theory, i.e. a theory living in half space, which has the advantage
that the full restriction of modular invariance can be exploited in the
construction of boundary states as pioneered by Cardy \cite{Cardy}. Since
this folding idea relies on the vanishing of either the reflection or
transmission, our considerations do in general not reduce to that set up,
even in the conformal limit. As was already pointed out in \cite{WA}, and as
can be seen directly from (\ref{1}) and (\ref{2}), in that case the
conductance is less interesting because it is either zero or perfect for
abelian theories. \ 

In section 2 we outline the procedure of how the defect scattering matrices
may be determined, since they are needed as input in both approaches. In
section 3 we newly formulate the defect TBA equations and use them to
determine the density distribution functions. We evaluate numerically the
Landauer formula (\ref{1}) for various defects and provide some analytical
approximations in certain regimes. In section 4 we propose a Kubo formula (%
\ref{2}) for a configuration in which an impurity is present and compute the
current-current two-point correlation functions occurring in there by means
of a form factor expansion. We find very good agreement between (\ref{1})
and (\ref{2}) for the complex free Fermion theory with various types of
defects. Our final conclusions and an outlook into open problems is provided
in section 5.

\section{Determining the defect scattering matrices}

\vspace{-0.2cm} \setcounter{equation}{0}

An essential input required in both non-perturbative methods which are
exploited to compute the conductance (\ref{1}) and (\ref{2}), that is the
TBA and the form factor bootstrap approach, respectively, is the knowledge
of the exact (defect) scattering matrix. It is one of the most intriguing
facts of two dimensional quantum field theories that these matrices can be
determined exactly to all orders in perturbation theory. In the following
section we will recall how much (little) of this approach can be carried
over to the situation when defects are present and compute explicitly the
transmission and reflection amplitudes for a variety of concrete defects.

\subsection{Defect Yang-Baxter equations}

\vspace{-0.2cm} \noindent A cornerstone in the context of integrable models
in 1+1 space-time dimensions are the Yang-Baxter equations \cite{YB}. They
can be derived most easily simply by exploiting the associativity of the
so-called Zamolodchikov-Faddeev (ZF) algebra \cite{FZ} and its extended
version which includes an additional generator representing a boundary \cite
{Chered,Skly,FK} or a defect \cite{DMS,CFG}. Indicating particle types by
Latin and degrees of freedom of the impurity by Greek letters, the
``braiding'' (exchange) relations of annihilation operators $Z_{i}(\theta )$
of a particle of type $i$ moving with rapidity $\theta $ and defect
operators $Z_{\alpha }$ in the state $\alpha $ can be written as 
\begin{eqnarray}
Z_{i}(\theta _{1})Z_{j}(\theta _{2}) &=&S_{ij}^{kl}(\theta _{1}-\theta
_{2})Z_{k}(\theta _{2})Z_{l}(\theta _{1}),  \label{Z0} \\
Z_{i}(\theta _{1})Z_{j}^{\dagger }(\theta _{2}) &=&S_{ij}^{kl}(\theta
_{1}-\theta _{2})Z_{k}^{\dagger }(\theta _{2})Z_{l}(\theta _{1})+2\pi \delta
(\theta _{1}-\theta _{2}),  \label{Z1/2} \\
Z_{i}(\theta )Z_{\alpha } &=&R_{i\alpha }^{j\beta }(\theta )Z_{j}(-\theta
)Z_{\beta }+T_{i\alpha }^{j\beta }(\theta )Z_{\beta }Z_{j}(\theta )\,,
\label{Z1} \\
Z_{\alpha }Z_{i}(\theta ) &=&\tilde{R}_{i\alpha }^{j\beta }(-\theta
)Z_{\beta }Z_{j}(-\theta )+\tilde{T}_{i\alpha }^{j\beta }(-\theta
)Z_{j}(\theta )Z_{\beta }.  \label{Z2}
\end{eqnarray}
The bulk scattering matrix is indicated by $S$, and the left/right
reflection and transmission amplitudes through the defect are denoted by $R/%
\tilde{R}$ and $T/\tilde{T}$, respectively. We employed Einstein's sum
convention, that is we assume sums over doubly occurring indices. We
suppress the explicit mentioning of the dependence of $Z_{\alpha }$ on the
position in space and assume for the time being that it is included in $%
\alpha $. For the treatment of a single defect this is not relevant, but it
will become important when we consider multiple defects. The same relations
hold when we replace the annihilation operators by the creation operators $%
Z_{i}^{\dagger }(\theta )$ with $R/\tilde{R}$, $T/\tilde{T}$ and $S$
replaced by their complex conjugates. The algebra (\ref{Z1})-(\ref{Z2}) can
be used to derive various relations amongst the scattering amplitudes. Using
extended ZF-algebra twice leads to the constraints 
\begin{eqnarray}
S_{ij}^{kl}(\theta )S_{kl}^{mn}(-\theta ) &=&\delta _{i}^{m}\delta _{j}^{n},
\label{U1} \\
R_{i\alpha }^{j\beta }(\theta )R_{j\beta }^{k\gamma }(-\theta )+T_{i\alpha
}^{j\beta }(\theta )\tilde{T}_{j\beta }^{k\gamma }(-\theta ) &=&\delta
_{i}^{k}\delta _{\alpha }^{\gamma },  \label{U2} \\
R_{i\alpha }^{j\beta }(\theta )T_{j\beta }^{k\gamma }(-\theta )+T_{i\alpha
}^{j\beta }(\theta )\tilde{R}_{j\beta }^{k\gamma }(-\theta ) &=&0\,.
\label{U3}
\end{eqnarray}
The same equations also hold after performing a parity transformation, that
is for $R\leftrightarrow \tilde{R}$ and $T\leftrightarrow \tilde{T}$ in (\ref
{U2})-(\ref{U3}). From the associativity of the extended ZF-algebra one
derives the equations \cite{Chered,Skly,FK,DMS,CFG} 
\begin{eqnarray}
S(\theta _{12})[\Bbb{I}\otimes R_{\alpha }^{\beta }(\theta _{1})]S(\hat{%
\theta}_{12})[\Bbb{I}\otimes R_{\beta }^{\gamma }(\theta _{2})]\!\!\!\!
&=&\!\!\!\![\Bbb{I}\otimes R_{\alpha }^{\beta }(\theta _{2})]S(\hat{\theta}%
_{12})[\Bbb{I}\otimes R_{\beta }^{\gamma }(\theta _{1})]S(\theta
_{12}),\,\,\,\,\,\,\,  \label{YBt1} \\
S(\theta _{12})[\Bbb{I}\otimes R_{\alpha }^{\beta }(\theta _{1})]S(\hat{%
\theta}_{12})[\Bbb{I}\otimes T_{\beta }^{\gamma }(\theta _{2})]\!\!\!\!
&=&\!\!\!\!R_{\beta }^{\gamma }(\theta _{1})\otimes T_{\alpha }^{\beta
}(\theta _{2}),  \label{YBt2} \\
S(\theta _{12})[T_{\alpha }^{\beta }(\theta _{2})\otimes T_{\beta }^{\gamma
}(\theta _{1})]\!\!\!\! &=&\!\!\!\![T_{\alpha }^{\beta }(\theta _{1})\otimes
T_{\beta }^{\gamma }(\theta _{2})]S(\theta _{12}),  \label{YBt3}
\end{eqnarray}
where we employed the convention $(A\otimes B)_{ij}^{kl}=A_{i}^{k}B_{j}^{l}$
for the tensor product and abbreviated the rapidity sum $\hat{\theta}%
_{12}=\theta _{1}+\theta _{2}$ and difference $\theta _{12}=\theta
_{1}-\theta _{2}$. Once again the same equations also hold for $%
R\leftrightarrow \tilde{R}$ and $T\leftrightarrow \tilde{T}$. Starting with
another initial asymptotic state one derives \cite{CFG} 
\begin{eqnarray}
R_{\alpha }^{\beta }(\theta _{1})\otimes \tilde{R}_{\beta }^{\gamma }(\theta
_{2}) &=&R_{\beta }^{\gamma }(\theta _{1})\otimes \tilde{R}_{\alpha }^{\beta
}(\theta _{2}),  \label{RR} \\
\lbrack T_{\alpha }^{\beta }(\theta _{2})\otimes \Bbb{I}]S(\hat{\theta}%
_{12})[\tilde{R}_{\beta }^{\gamma }(\theta _{1})\otimes \Bbb{I}]S(\theta
_{12}) &=&T_{\beta }^{\gamma }(\theta _{2})\otimes \tilde{R}_{\alpha
}^{\beta }(\theta _{1}),  \label{TR} \\
\lbrack \Bbb{I}\otimes \tilde{T}_{\alpha }^{\beta }(\theta _{2})]S(\hat{%
\theta}_{12})[\Bbb{I}\otimes R_{\beta }^{\gamma }(\theta _{1})]S(\theta
_{12}) &=&R_{\alpha }^{\beta }(\theta _{1})\otimes \tilde{T}_{\beta
}^{\gamma }(\theta _{2}),  \label{RT} \\
\lbrack T_{\alpha }^{\beta }(\theta _{1})\otimes \Bbb{I}]S(\hat{\theta}%
_{12})[\tilde{T}_{\beta }^{\gamma }(\theta _{2})\otimes \Bbb{I}] &=&[\Bbb{I}%
\otimes \tilde{T}_{\alpha }^{\beta }(\theta _{2})]S(\hat{\theta}_{12})[\Bbb{I%
}\otimes T_{\beta }^{\gamma }(\theta _{1})].
\end{eqnarray}

\noindent On the basis of the equations (\ref{YBt1})-(\ref{YBt3}), it was
shown in \cite{DMS}, for the abelian case without defect degrees of freedom,
that one can not have reflection and transmission simultaneously. In \cite
{CFG} this result was extended to the non-abelian parity breaking case and
it was proven that for the simultaneous occurrence of reflection and
transmission the scattering matrix has to be rapidity independent and of the
form 
\begin{equation}
S(\theta )\Bbb{=P}\sigma \,,  \label{S}
\end{equation}
with $\Bbb{P}$ being a permutation operator and $\sigma $ a constant matrix.
When assuming in addition that $\sigma $ is a diagonal matrix with the
property $\sigma _{ij}\sigma _{ji}=1$, the free Fermion ($\sigma
_{ij}=\sigma _{ji}=-1$), free Boson ($\sigma _{ij}=\sigma _{ji}=1$) and also
the Federbush model \cite{Feder} and the generalized coupled Federbush
models \cite{Fform} are solutions to (\ref{S}).

\noindent As a further set of consistency equations, which serve for the
determination of the defect scattering matrix, we report the crossing
relations, which are as usual less obvious to justify. In analogy to the
relations which have to hold for the bulk scattering matrix $S_{ij}(\theta
)=S_{i\bar{\jmath}}(i\pi -\theta )=S_{ji}^{\ast }(-\theta )$, ($\bar{\jmath}$
is the anti-particle of $j$ and $\ast $ denotes the complex conjugation) we
deduce from (\ref{Z1})-(\ref{Z2}) the crossing-hermiticity relations 
\begin{eqnarray}
R_{_{\bar{\jmath}}}^{\alpha }(\theta ) &=&\tilde{R}_{_{\bar{\jmath}%
}}^{\alpha }(-\theta )^{\ast }=S_{j\bar{\jmath}}(2\theta )\tilde{R}%
_{j}^{\alpha }(i\pi -\theta )\,,  \label{c1} \\
T_{_{\bar{\jmath}}}^{\alpha }(\theta ) &=&\tilde{T}_{_{\bar{\jmath}%
}}^{\alpha }(-\theta )^{\ast }=\tilde{T}_{j}^{\alpha }(i\pi -\theta )\,.
\label{c2}
\end{eqnarray}
The first equalities follow when taking $Z_{i}^{\dagger }(\theta )^{\ast
}=Z_{i}(\theta )$ and $Z_{\alpha }=Z_{\alpha }^{\dagger }$. The latter
relations in (\ref{c2}) simply result by considering the relations for $S$
while letting one of the particles freeze, i.e., setting its rapidity to
zero, and viewing it as a defect. Relations (\ref{c1}) are obtainable in a
similar fashion as the interpretation put forward in \cite{GZ,FK}. Our
equations (\ref{c1}) and (\ref{c2}) disagree slightly from the crossing
relations in \cite{GZ,FK,DMS,Konik}, which is due to the fact that when
parity is broken real analyticity is replaced by Hermitian analyticity \cite
{HERMAN}. Later on in our example, this will also be reflected in the
representation of the free Fermion field (\ref{free}), being Dirac rather
than Majorana. There is of course no consequence of this choice of
conventions on the physics, since the ambiguity just exploits the fact that
only the moduli of these amplitudes are observable.

Similar as for the bulk scattering matrices an additional powerful
constraint results from the singularity structure of the defect scattering
amplitudes. In \cite{CFG} it was shown that the defect does not admit any
excited state once one demands a simultaneous occurrence of reflection and
transmission. Supposing that the defect scattering matrices have a pole on
the imaginary axis at $i\theta _{0}\in i\Bbb{R}$, the corresponding residues
are therefore constraint as 
\begin{equation}
\limfunc{Res}_{\theta \rightarrow i\theta _{0}}R_{j}^{\alpha }(\theta )=%
\limfunc{Res}_{\theta \rightarrow i\theta _{0}}\tilde{R}_{j}^{\alpha
}(\theta )=\limfunc{Res}_{\theta \rightarrow i\theta _{0}}T_{j}^{\alpha
}(\theta )=\limfunc{Res}_{\theta \rightarrow i\theta _{0}}\tilde{T}%
_{j}^{\alpha }(\theta )\left\{ 
\begin{array}{l}
<0\quad \text{for \thinspace \thinspace }\theta _{0}\in (0,\pi ) \\ 
>0\quad \text{for \thinspace \thinspace }\theta _{0}\notin (0,\pi )
\end{array}
\right. \,.  \label{res}
\end{equation}
The intervals $(0,\pi )$ are understood to be $\func{mod}$\thinspace $2\pi $%
. Hence, there is no pole with positive residue in the physical sheet.

\subsection{Multiple defects}

\vspace{-0.2cm}

\noindent Assuming that we have determined the defect scattering matrices $R/%
\tilde{R}$ and $T/\tilde{T}$ for a single defect, for instance by solving
the consistency equations in the previous subsection, it is straightforward
to use them in order to compute the related quantities for several defects.
This type of situation is of interest since, unlike for a single defect, it
leads to the occurrence of resonance phenomena and when the number of
defects tends to infinity even to band structures. Let us therefore commence
by exploiting the extended ZF-algebra (\ref{Z1})-(\ref{Z2}) for a double
defect. For the reasons mentioned in the introduction we are interested in
the situation when $R/\tilde{R}$ and $T/\tilde{T}$ are simultaneously
non-vanishing, and in the light of the result (\ref{S}), we shall therefore
focus on the diagonal case from now onwards. We compute 
\begin{eqnarray}
Z_{i}(\theta )Z_{\alpha }Z_{\beta } &=&R_{i}^{\alpha \beta }(\theta
)Z_{i}(-\theta )Z_{\alpha }Z_{\beta }+T_{i}^{\alpha \beta }(\theta
)Z_{\alpha }Z_{\beta }Z_{i}(\theta )\,, \\
Z_{\alpha }Z_{\beta }Z_{i}(\theta ) &=&\tilde{R}_{i}^{\alpha \beta }(-\theta
)Z_{\alpha }Z_{\beta }Z_{i}(-\theta )+\tilde{T}_{i}^{\alpha \beta }(-\theta
)Z_{i}(\theta )Z_{\alpha }Z_{\beta },
\end{eqnarray}
where we have now introduced overall transmission and reflection amplitudes
corresponding to two defects 
\begin{eqnarray}
T_{i}^{\alpha \beta }(\theta ) &=&\frac{T_{i}^{\alpha }(\theta )T_{i}^{\beta
}(\theta )}{1-R_{i}^{\beta }(\theta )\tilde{R}_{i}^{\alpha }(\theta )}%
,\qquad R_{i}^{\alpha \beta }(\theta )=R_{i}^{\alpha }(\theta )+\frac{%
R_{i}^{\beta }(\theta )T_{i}^{\alpha }(\theta )\tilde{T}_{i}^{\alpha
}(\theta )}{1-R_{i}^{\beta }(\theta )\tilde{R}_{i}^{\alpha }(\theta )},
\label{tr} \\
\tilde{T}_{i}^{\alpha \beta }(\theta ) &=&\frac{\tilde{T}_{i}^{\alpha
}(\theta )\tilde{T}_{i}^{\beta }(\theta )}{1-R_{i}^{\beta }(\theta )\tilde{R}%
_{i}^{\alpha }(\theta )},\qquad \tilde{R}_{i}^{\alpha \beta }(\theta )=%
\tilde{R}_{i}^{\beta }(\theta )+\frac{R_{i}^{\alpha }(\theta )T_{i}^{\beta
}(\theta )\tilde{T}_{i}^{\beta }(\theta )}{1-R_{i}^{\beta }(\theta )\tilde{R}%
_{i}^{\alpha }(\theta )}.  \label{tr2}
\end{eqnarray}
The term $[1-R_{i}^{\beta }(\theta )\tilde{R}_{i}^{\alpha }(\theta
)]^{-1}=\sum_{n=1}^{\infty }(R_{i}^{\beta }(\theta )\tilde{R}_{i}^{\alpha
}(\theta ))^{n}$ results from the infinite number of reflections which we
have in-between the two defects, well known from Fabry-Perot type devices of
classical and quantum optics. For the case $T=\tilde{T},R=\tilde{R}$ the
expressions (\ref{tr}) and (\ref{tr2}) coincide with the formulae proposed
in \cite{Konik2}. When absorbing the space dependent phase factor into the
defect matrices, the explicit example presented in \cite{DMS} for the free
fermion perturbed with the energy operator agree almost for $T=\tilde{T},R=%
\tilde{R}$ \ with the general formulae (\ref{tr}). They disagree in the
sense that the equality of $R_{i}^{\alpha \beta }(\theta )$ and $\tilde{R}%
_{i}^{\alpha \beta }(\theta )$ does not hold for generic $\alpha ,\beta $ as
stated in \cite{DMS}.

It is now straightforward to extend the expressions to an arbitrary number
of defects, say $n$, in a recursive manner 
\begin{eqnarray}
T_{i}^{\mathbf{\alpha }}(\theta ) &=&\frac{T_{i}^{\alpha _{1}\ldots \alpha
_{k}}(\theta )T_{i}^{\alpha _{k+1}\ldots \alpha _{n}}(\theta )}{1-\tilde{R}%
_{i}^{\alpha _{1}\ldots \alpha _{k}}(\theta )R_{i}^{\alpha _{k+1}\ldots
\alpha _{n}}(\theta )},\qquad \,\,\qquad \qquad \qquad
\,\,\,\,\,\,\,\,\,\,\,1<k<n\,,  \label{ttr} \\
R_{i}^{\mathbf{\alpha }}(\theta ) &=&R_{i}^{\alpha _{1}\ldots \alpha
_{k}}(\theta )+\frac{R_{i}^{\alpha _{k+1}\ldots \alpha _{n}}(\theta
)T_{i}^{\alpha _{1}\ldots \alpha _{k}}(\theta )\tilde{T}_{i}^{\alpha
_{1}\ldots \alpha _{k}}(\theta )}{1-\tilde{R}_{i}^{\alpha _{1}\ldots \alpha
_{k}}(\theta )R_{i}^{\alpha _{k+1}\ldots \alpha _{n}}(\theta )},\quad
\,\,1<k<n\,.\,\,  \label{ttr2}
\end{eqnarray}
For convenience we encoded here the defect degrees of freedom into the
vector $\mathbf{\alpha =}\{\alpha _{1},\cdots ,\alpha _{n}\}$. Similar
expressions also hold for $\tilde{T}_{i}^{\mathbf{\alpha }}(\theta )=\tilde{T%
}_{i}^{\alpha _{1}\ldots \alpha _{n}}(\theta )$ and $\tilde{R}_{i}^{\mathbf{%
\alpha }}(\theta )=\tilde{R}_{i}^{\alpha _{1}\ldots \alpha _{n}}(\theta )$.
It is clear that from the knowledge of the single defect amplitudes we are
now in the position to compute the corresponding quantities for multiple
defects just by nesting successively the expressions (\ref{ttr}) and (\ref
{ttr2}) for increasing values of $n$ into each other. Nonetheless, in
general one does not succeed to provide simple analytical expressions for\ $n
$-defect amplitudes and a different description is useful.

Alternatively, we can define, in analogy to standard quantum mechanical
methods (see e.g. \cite{CT,Merz}), a transmission matrix which takes the
particle from one side of the defect to the other. From the braiding
relations (\ref{Z1}) and (\ref{Z2}), we obtain with the help of the
unitarity relations (\ref{U2}) and (\ref{U3}) 
\begin{equation}
\binom{Z_{\alpha _{1}}\ldots Z_{\alpha _{n}}Z_{i}(\theta )}{Z_{\alpha
_{1}}\ldots Z_{\alpha _{n}}Z_{i}(-\theta )}=\left( \prod_{k=1}^{n}\mathcal{M}%
_{\alpha _{k}}^{i}(\theta )\right) \binom{Z_{i}(\theta )Z_{\alpha
_{1}}\ldots Z_{\alpha _{n}}}{Z_{i}(-\theta )Z_{\alpha _{1}}\ldots Z_{\alpha
_{n}}}\,,  \label{mult}
\end{equation}
with 
\begin{equation}
\mathcal{M}_{\alpha _{k}}^{i}(\theta )=\left( 
\begin{array}{cc}
T_{i}^{\alpha _{k}}(\theta )^{-1} & -R_{i}^{\alpha _{k}}(\theta
)T_{i}^{\alpha _{k}}(\theta )^{-1} \\ 
-R_{i}^{\alpha _{k}}(-\theta )T_{i}^{\alpha _{k}}(-\theta )^{-1} & 
T_{i}^{\alpha _{k}}(-\theta )^{-1}
\end{array}
\right) \,.
\end{equation}
This means alternatively to the recursive way (\ref{ttr}) and (\ref{ttr2}),
we can also compute the multi-defect transmission and reflection amplitudes
as 
\begin{equation}
T_{i}^{\mathbf{\alpha }}(\theta )=\left( \prod_{k=1}^{n}\mathcal{M}_{\alpha
_{k}}^{i}(\theta )\right) _{11}^{-1},\,\,\,\quad R_{i}^{\mathbf{\alpha }%
}(\theta )=-\left( \prod_{k=1}^{n}\mathcal{M}_{\alpha _{k}}^{i}(\theta
)\right) _{12}\left( \prod_{k=1}^{n}\mathcal{M}_{\alpha _{k}}^{i}(\theta
)\right) _{11}^{-1}.  \label{ttr3}
\end{equation}
One may convince oneself that this formulation is indeed the same as (\ref
{ttr}) and (\ref{ttr2}). It has, however, the virtue that it allows for a
more elegant computation of the band structures. In particular, it is most
suitable for numerical computations, since it just involves matrix
multiplications rather than recurrence operations.

Let us now consider the case in which all the defects are of the same type $%
\alpha $, equidistantly separated by an amount $y$ and send $n\rightarrow
\infty $. First of all we have to include now explicitly the dependence of
the defect on its position into the discussion. We assume 
\begin{equation}
\prod_{l=1}^{n}\mathcal{M}_{\alpha }^{i}(x=ly)=\prod_{l=1}^{n}\left[ Q_{y}%
\mathcal{M}_{\alpha }^{i}(x=0)\right] ^{l}Q_{ny}^{-1}\,,\quad Q_{y}=\left( 
\begin{array}{cc}
e^{iky} & 0 \\ 
0 & e^{-iky}
\end{array}
\right) \,,  \label{yx}
\end{equation}
where $k$ corresponds to the wavevector of the lattice. Taking then $%
n\rightarrow \infty $ this accommodates Bloch's theorem (e.g., \cite{CT})
for the relativistic set-up. The simple requirement, that the product of
transmission matrices $\lim_{n\rightarrow \infty }\prod_{l=1}^{n}\mathcal{M}%
_{\alpha }^{i}(x=ly)$ remains finite, leads now in the usual way to a
restriction for the allowed energies, that is to band structures. To see
when this is the case we can exploit the r.h.s. of the first equation in (%
\ref{yx}) and diagonalize the matrix $Q_{y}\mathcal{M}_{\alpha }^{i}(x=0)$.
Then it is clear that the limit $n\rightarrow \infty $ only remains finite
when the eigenvalues of this matrix are not real 
\begin{equation}
\lambda ^{i,\alpha }\notin \Bbb{R\,}.  \label{unreal}
\end{equation}
The eigenvalues are computed to 
\begin{eqnarray}
\lambda _{1,2}^{i,\alpha } &=&\chi _{i}^{\alpha }(\theta )\pm \sqrt{\chi
_{i}^{\alpha }(\theta )^{2}-\tilde{T}_{i}^{\alpha }(-\theta )/T_{i}^{\alpha
}(-\theta )},\quad   \label{lam} \\
\chi _{i}^{\alpha }(\theta ) &=&\frac{[e^{iky}T_{i}^{\alpha }(\theta
)^{-1}+e^{-iky}(\tilde{T}_{i}^{\alpha }(\theta )^{\ast })^{-1}]}{2}.\,\,
\label{chi}
\end{eqnarray}
In the parity invariant case the criterium (\ref{unreal}) becomes simpler.
From (\ref{lam}) and (\ref{chi}) follows in that case that the allowed
energies in the infinite lattice have to respect 
\begin{equation}
\chi _{i}^{\alpha }(\theta )=\func{Re}[e^{iky}T_{i}^{\alpha }(\theta )^{-1}%
]<1,\qquad \text{for }T=\tilde{T}\,\,.  \label{bands}
\end{equation}
In other words particles are only allowed to travel in the system with
rapidities for which the inequality (\ref{bands}) holds. In conclusion, this
means the determination of the transmission amplitudes for a single defect
is sufficient to determine multiple defects and the energy band structure.
Let us illustrate the working of this general formulae with a concrete
example.

\subsection{Free Fermion with defects}

\vspace{-0.2cm}

\noindent The continuous version of the 1+1 dimensional free Fermion (Ising
model) with a line of defect was first treated in \cite{Cabra}. Thereafter
it has also been considered in \cite{GZ,DMS} and \cite{Konik} from a
different point of view. In \cite{Cabra,GZ,DMS} the defect line has the form
of the energy operator and in \cite{Konik} also a perturbation in form of a
single Fermion has been considered. In this manuscript we want to enlarge
the class of perturbations having in mind to obtain various different kinds
of structural and physical behaviours.

Let us consider the Lagrangian density for a complex free Fermion $\psi $
with $\ell $ defects\footnote{%
Throughout the paper we use the following conventions: 
\begin{eqnarray*}
x^{\mu } &=&(x^{0},x^{1}),\qquad p^{\mu }=(m\cosh \theta ,m\sinh \theta
),\quad g^{00}=-g^{11}=\varepsilon ^{01}=-\varepsilon ^{10}=1, \\
\gamma ^{0} &=&\left( 
\begin{array}{cc}
0 & 1 \\ 
1 & 0
\end{array}
\right) ,\quad \quad \gamma ^{1}=\left( 
\begin{array}{cc}
0 & 1 \\ 
-1 & 0
\end{array}
\right) ,\quad \gamma ^{5}=\gamma ^{0}\gamma ^{1},\quad \quad \psi _{\alpha
}=\left( 
\begin{array}{c}
\psi _{\alpha }^{(1)} \\ 
\psi _{\alpha }^{(2)}
\end{array}
\right) ,\quad \bar{\psi}_{\alpha }=\psi _{\alpha }^{\dagger }\gamma ^{0}\,.
\end{eqnarray*}
} 
\begin{equation}
\mathcal{L}=\bar{\psi}(i\gamma ^{\mu }\partial _{\mu }-m)\psi
\,+\sum\limits_{n=0}^{\ell -1}\delta (x-x_{n})\mathcal{D}^{\alpha _{n}}(\bar{%
\psi},\psi )\,,
\end{equation}
where we describe the defect by the functions $\mathcal{D}^{\alpha _{n}}(%
\bar{\psi},\psi ),$ which we assume to be linear in the Fermi fields $\bar{%
\psi}$ and $\psi $. In the following we will restrict ourselves mainly to
the case of equidistantly distributed defects of the same type, i.e. $%
x_{n}=ny$ and $\mathcal{D}^{\alpha _{n}}(\bar{\psi},\psi )=\mathcal{D}(\bar{%
\psi},\psi )$ for $n\in \{0,\ell -1\}$.

\subsubsection{Transmission and reflection amplitudes}

\vspace{-0.2cm}

\noindent Unfortunately, it follows from the arguments outlined in section
2.1, that when one is seeking a situation with simultaneously occurring
reflection and transmission the constraining equations for diagonal bulk
scattering matrices reduce simply to unitarity and crossing. These equations
are, however, not restrictive enough by themselves to fix $R/\tilde{R}$ and $%
T/\tilde{T}$ and therefore one has to resort to alternative arguments. For
instance one may proceed in analogy to standard quantum mechanical potential
scattering theory (see also \cite{GZ,DMS,Konik}) and construct the
amplitudes by adequate matching conditions on the field. We consider now a
single defect at the origin which suffices, since multiple defect amplitudes
can be constructed from the single defect ones, according to the arguments
of the previous section. We decompose the fields of the bulk theory as $\psi
(x)=\Theta (x)$ $\psi _{+}(x)+\Theta (-x)$ $\psi _{-}(x)$, with $\Theta (x)$
being the Heavyside step function, and substitute this ansatz into the
equations of motion. This way we obtain the constraints 
\begin{equation}
i\gamma ^{1}(\psi _{+}(x)-\psi _{-}(x))|_{x=0}=\left. \frac{\partial 
\mathcal{D}(\bar{\psi}(x),\psi (x))}{\partial \bar{\psi}(x)}\right|
_{x=0}\,\,.  \label{bcon}
\end{equation}
Using here for the left ($-$) and right ($+$) parts of $\psi $ the Fourier
decomposition of the free field 
\begin{equation}
\psi _{j}(x)=\int \frac{dp_{j}^{1}}{\sqrt{4\pi }p_{j}^{0}}\left(
a_{j}(p)u_{j}(p)e^{-ip_{j}\cdot x}+a_{_{\bar{\jmath}}}^{\dagger
}(p)v_{j}(p)e^{ip_{j}\cdot x}\right) \,,\qquad  \label{free}
\end{equation}
with $\sqrt{m_{j}^{2}+p_{j}^{2}}=p_{j}^{0}$ and the Weyl spinors 
\begin{equation}
u_{j}(p)=\sqrt{\frac{m_{j}}{2}}\left( 
\begin{array}{c}
e^{-\theta /2} \\ 
e^{\theta /2}
\end{array}
\right) \quad \quad \text{and\qquad }v_{j}(p)=i\sqrt{\frac{m_{j}}{2}}\left( 
\begin{array}{c}
e^{-\theta /2} \\ 
-e^{\theta /2}
\end{array}
\right) \,\,,  \label{WS}
\end{equation}
we can substitute them into the constraint (\ref{bcon}). Treating the
equations obtained in this manner componentwise, stripping off the
integrals, we can bring them thereafter into the form 
\begin{eqnarray}
\left( 
\begin{array}{c}
a_{_{\bar{\jmath}},-}^{\dagger }(\theta ) \\ 
a_{_{\bar{\jmath}},+}^{\dagger }(-\theta )
\end{array}
\right) &=&\left( 
\begin{array}{cc}
R_{_{\bar{\jmath}}}(\theta )^{\ast } & T_{_{\bar{\jmath}}}(\theta )^{\ast }
\\ 
\tilde{T}_{_{\bar{\jmath}}}(\theta )^{\ast } & \tilde{R}_{_{\bar{\jmath}%
}}(\theta )^{\ast }
\end{array}
\right) \left( 
\begin{array}{c}
a_{_{\bar{\jmath}},-}^{\dagger }(-\theta ) \\ 
a_{_{\bar{\jmath}},+}^{\dagger }(\theta )
\end{array}
\right) \,,  \label{m1} \\
\left( 
\begin{array}{c}
a_{j,-}(\theta ) \\ 
a_{j,+}(-\theta )
\end{array}
\right) &=&\left( 
\begin{array}{cc}
R_{_{j}}(\theta ) & T_{_{j}}(\theta ) \\ 
\tilde{T}_{_{j}}(\theta ) & \tilde{R}_{j}(\theta )
\end{array}
\right) \left( 
\begin{array}{c}
a_{j,-}(-\theta ) \\ 
a_{j,+}(\theta )
\end{array}
\right) \,.  \label{m2}
\end{eqnarray}
The creation and annihilation operators $a_{i}(\theta ),\,a_{i}^{\dagger
}(\theta )$ play in (\ref{Z0}) and (\ref{Z1/2}) the role of the ZF-algebra
generators in view of the usual fermionic anti-commutation relations $%
\{a_{i}(\theta _{1}),a_{j}(\theta _{2})\}=0$, $\{a_{i}(\theta
_{1}),a_{j}^{\dagger }(\theta _{2})\}=2\pi \delta _{ij}\delta (\theta _{12})$%
. When including the defect operator in the equations (\ref{m1}) and (\ref
{m2}), on the right/left for $-/+$-subscript, they acquire precisely the
form of the extended ZF-algebra (\ref{Z1})-(\ref{Z2}), such that one can
read off directly the reflection and transmission amplitudes. One may
convince oneself that the expressions found this way indeed satisfy the
consistency equations like crossing (\ref{c1}), (\ref{c2}), unitarity (\ref
{U2}), (\ref{U3}) and respect (\ref{res}). In order to find the explicit
expressions, we have to consider some concrete defects. Let us first
concentrate on the energy perturbation.

\subsubsection{The energy operator defect ${\mathcal{D}^{\protect\alpha } (%
\bar{\protect\psi},\protect\psi )=g\bar{\protect\psi}\protect\psi} $}

\vspace{-0.2cm}

\noindent The defect $\mathcal{D}^{\alpha }(\bar{\psi},\psi )=g\bar{\psi}%
\psi $ has received already some amount of consideration, for the reason
that it possesses a well studied \cite{disIs} discrete counterpart. Taking
the continuum limit of these lattice models the defect term in there
acquires the form of the energy operator $\varepsilon (x)=g\bar{\psi}\psi
(x) $, with $g$ being a coupling constant. According to (\ref{bcon}), (\ref
{m1}) and (\ref{m2}) we compute 
\begin{eqnarray}
\tilde{R}_{j}^{\alpha }(\theta ,y) &=&R_{\bar{\jmath}}^{\alpha }(\theta
,y)=R_{j}^{\alpha }(\theta ,-y)=\tilde{R}_{\bar{\jmath}}^{\alpha }(\theta
,-y)=\frac{\sin B\cosh \theta }{i\sinh \theta -\sin B}e^{2iym\sinh \theta
}\,,  \label{ren} \\
T_{j}^{\alpha }(\theta ) &=&\tilde{T}_{j}^{\alpha }(\theta )=T_{\bar{\jmath}%
}^{\alpha }(\theta )=\tilde{T}_{\bar{\jmath}}^{\alpha }(\theta )=\frac{\cos
B\sinh \theta }{\sinh \theta +i\sin B}\,,  \label{ten}
\end{eqnarray}
where we used a common and convenient parameterization in this context%
\footnote{%
This is suggestive since many bulk theories admit such a relation between
the bare and effective coupling. One may equate some combinations of $R$ and 
$T$ with some well known bulk scattering matrices. For instance, we identify
the sinh-Gordon S-matrix $S_{SG}(\theta ,B_{SG})=T_{j}(\theta ,B\pi
/2)/T_{j}(-\theta ,B\pi /2)$, with the indicated relation amongst the
effective defect coupling constants.} 
\begin{equation}
\sin B=-\frac{4g}{4+g^{2}},\qquad -\frac{\pi }{2}\leq B\leq 0\,.
\label{coup}
\end{equation}
Note, that there is no explicit $y$-dependence in $T/\tilde{T}$ and that (%
\ref{ren})-(\ref{ten}) satisfy the ``unitarity'' relations (\ref{U2})-(\ref
{U3}) and the crossing-hermiticity relations (\ref{c1})-(\ref{c2}) when the
defect is situated at the origin. The residues are constrained as in (\ref
{res}). The expressions $R_{j}^{\alpha }(\theta ,B)$ and $T_{j}^{\alpha
}(\theta ,B)$ coincide with the solutions found in \cite{DMS}, which,
however, in general does not correspond to taking our particles simply to be
self-conjugated, since we use Dirac Fermions. Having obtained these
amplitudes, we can easily compute the corresponding quantities associated to
multiple defects by means of (\ref{ttr}), (\ref{ttr2}) or (\ref{ttr3}). The
explicit formulae are obvious and since they are quite cumbersome we will
not report them here. Instead, we will depict them as functions of $\cosh
\theta $ in figure 1 for various parameters in order to emphasize some of
their characteristics.\bigskip

\begin{center}
\includegraphics[width=15.0cm,height=11.14cm]{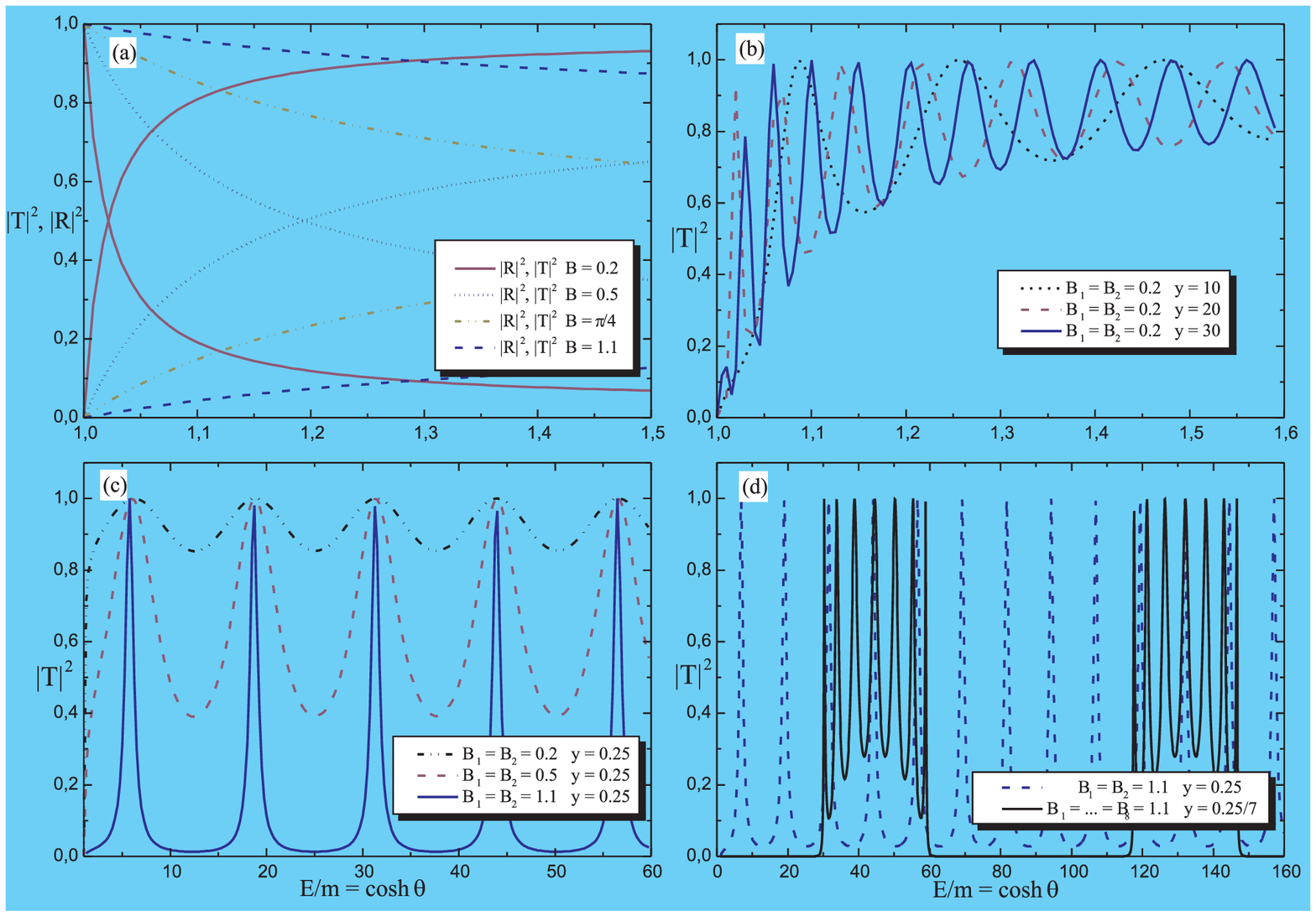}
\end{center}

\noindent {\small Figure 1: (a) Single defect with varying coupling
constant. $|T|^{2}$ and $|R|^{2}$ correspond to curves starting at 0 and 1
of the same line type, respectively. (b) Double defect with varying distance 
}$y${\small . (c) Double defect with varying effective coupling constant }$B$%
{\small . (d) Double defect $\equiv $ \ dotted line, eight defects $\equiv $
solid line.}

Part (a) of figure 1 confirms the unitarity relation (\ref{U2}) where we
used $R_{j}^{\ast }(\theta ,B)=R_{j}(-\theta ,B)$ and $T_{j}^{\ast }(\theta
,B)=T_{j}(-\theta ,B)$. Part (b) and (c) show the typical resonances of a
double defect, which become stretched out and pronounced with respect to the
energy when the distance becomes smaller and the coupling constant
increases, respectively. Part (d) exhibits a general feature which extends
to an even number of higher multiple defects, say $2n$, when keeping the
distance $y$ between the two most separated defects fixed: The resonances
accumulate at the position around the ($2n-1$)-th resonances of the double
defect. For increasing $n$ they become very dense in that region such that
one may speak of energy bands.

It is interesting to compare these bands with those obtained from the
criterium (\ref{bands}), which translates in this case into 
\begin{equation}
\sinh \theta (\cos ky-\cos B)<\sin ky\sin B,\qquad k=m\sinh \theta \,.
\end{equation}
Figure 1 part (d) shows that when taking $2n$ defects separated by a
distance $y/(2n-1)$ one obtains for large $n$ an energy spectrum which
resembles a band structure. Analyzing instead the function $\chi
_{i}^{\alpha }(\theta )$ in (\ref{chi}) we obtain the same band structure
from the criterium (\ref{bands}). The two computations show that the
positions as well as the width of the bands in the two figures 1(d) and 2
coincide quite well. Remarkably, even for the double defect the criterium (%
\ref{chi}) yields energy regions, see figure 2(b), which are in good
agreement with the exact computation as presented in figure 1(d).

\medskip

\begin{center}
\includegraphics[width=15.cm,height=5.7cm]{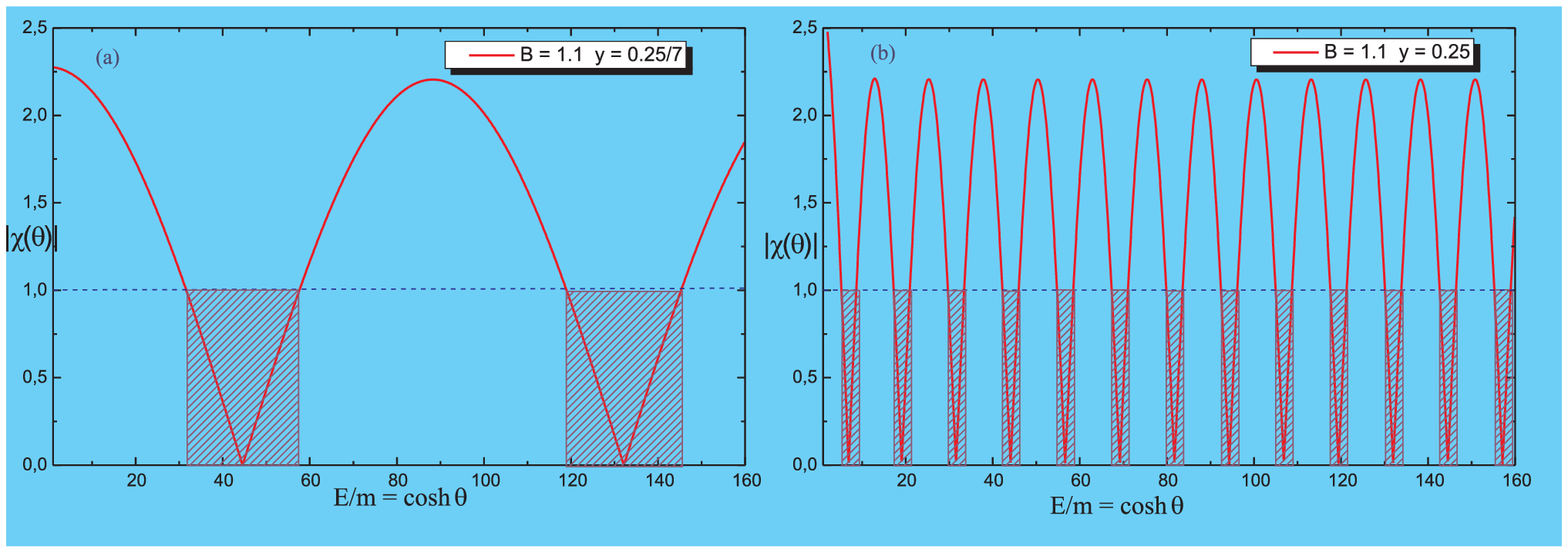}
\end{center}

\noindent {\small Figure 2: Band structures according to the criterium (\ref
{bands}). The non-shaded regions are forbidden. (a) eight defects with $%
B=B_{1}=\ldots =B_{8}=1.1$ equidistant by $y=0.25/7$ (b) double defect with $%
B=B_{1}=B_{2}=1.1$ distanced by $y=0.25.$ }

Very often we will not be able to perform certain computations analytically,
but instead we can carry them out in the massless limit. The prescription
for taking this limit was originally introduced in \cite{massless}. It
consists of replacing in every rapidity dependent expression $\theta $ by $%
\theta \pm \sigma $, where an additional auxiliary parameter $\sigma $ has
been introduced. Thereafter one should take the limit $\sigma \rightarrow
\infty $, $m\rightarrow 0$ while keeping the quantity $\hat{m}=m/2\exp
(\sigma )$ finite. For instance, carrying out this prescription for the
momentum yields $p_{\pm }=\pm \hat{m}\exp (\pm \theta )$, such that one may
view the model as splitted into its two chiral sectors and one can speak
naturally of left (L) and right (R) movers. In this way the expressions (\ref
{ren})-(\ref{ten}) become 
\begin{equation}
R_{j,L/R}^{\alpha }(\theta ,B)=\pm i\sin B\,e^{\pm 2iy_{\alpha }\hat{m}%
e^{\theta }}\,\,\,\,\,\quad \text{and}\,\,\,\,\quad \,\,\,T_{j,L/R}^{\alpha
}(\theta ,B)=\cos B.  \label{LR0}
\end{equation}
Similarly we compute the expression involving two and four defects for later
purposes 
\begin{eqnarray}
T_{j,L/R}^{\alpha _{1}\alpha _{2}}(\theta ,B) &=&\tilde{T}_{j,L/R}^{\alpha
_{1}\alpha _{2}}(\theta ,B)=\frac{\cos ^{2}B}{1+\sin ^{2}B\exp [\mp 2i\hat{m}%
(y_{\alpha _{1}}-y_{\alpha _{2}})e^{\theta }]}\,,  \label{LR2} \\
R_{j,L/R}^{\alpha _{1}\alpha _{2}}(\theta ,B) &=&\pm i\sin
B\,e^{-2iy_{\alpha _{1}}\hat{m}e^{\theta }}\pm \frac{i\sin B\cos
^{2}Be^{-2iy_{\alpha _{2}}\hat{m}e^{\theta }}}{1+\sin ^{2}B\exp [\mp 2i\hat{m%
}(y_{\alpha _{1}}-y_{\alpha _{2}})e^{\theta }]}\,, \\
\tilde{R}_{j,L/R}^{\alpha _{1}\alpha _{2}}(\theta ,B) &=&\pm i\sin
B\,e^{2iy_{\alpha _{2}}\hat{m}e^{\theta }}\pm \frac{i\sin B\cos
^{2}Be^{-2iy_{\alpha _{1}}\hat{m}e^{\theta }}}{1+\sin ^{2}B\exp [\mp 2i\hat{m%
}(y_{\alpha _{1}}-y_{\alpha _{2}})e^{\theta }]}\,,\qquad \\
T_{j,L/R}^{\alpha _{1}\alpha _{2}\alpha _{3}\alpha _{4}}(\theta ,B) &=&%
\tilde{T}_{j,L/R}^{\alpha _{1}\alpha _{2}\alpha _{3}\alpha _{4}}(\theta ,B)=%
\frac{T_{j,L/R}^{\alpha _{1}\alpha _{2}}(\theta ,B)T_{j,L/R}^{\alpha
_{3}\alpha _{4}}(\theta ,B)}{1-\tilde{R}_{j,L/R}^{\alpha _{1}\alpha
_{2}}(\theta ,B)R_{j,L/R}^{\alpha _{3}\alpha _{4}}(\theta ,B)}\,.
\label{LR5}
\end{eqnarray}
The remaining amplitudes can be obtained analogously. The expressions of
physical quantities, e.g., the conductance, in the massless limit should not
depend on the parameter $\hat{m}$, such that the amplitudes (\ref{LR0})-(\ref
{LR5}) should in fact always appear in combination with other functions in
order to make the prescription meaningful.

Having discussed this type of defect in some detail we will now compute $R/%
\tilde{R}$ and $T/\tilde{T}$ for various other defects in order to
illustrate several types of physical behaviours.

\subsubsection{Transparent defects, $\mathcal{D}^{0}(\bar{\protect\psi},%
\protect\psi )=0$, $\mathcal{D}^{\protect\beta }(\bar{\protect\psi},\protect%
\psi )=g\bar{\protect\psi}\protect\gamma ^{1}\protect\psi $}

\vspace{-0.2cm} The examples which can be handled most easily in later
considerations are defects which behave physically as if they were
transparent ones, i.e., as $|T^{\mathbf{\alpha }}|=1$. Note that this does
not necessarily mean the absence of the defect. For instance considering the
defect $\mathcal{D}^{\beta }(\bar{\psi},\psi )=g\bar{\psi}\gamma ^{1}\psi $,
we compute with the method outlined above 
\begin{eqnarray}
R_{j}^{\beta }(\theta ,B)\!\! &=&\!\!\tilde{R}_{j}^{\beta }(\theta ,B)=R_{%
\bar{\jmath}}^{\beta }(\theta ,B)=\tilde{R}_{\bar{\jmath}}^{\beta }(\theta
,B)=0\,,  \label{Rb} \\
T_{j}^{\beta }(\theta ,-B)\!\! &=&\!\!T_{\bar{\jmath}}^{\beta }(\theta ,B)=%
\tilde{T}_{\bar{\jmath}}^{\beta }(\theta ,-B)=\tilde{T}_{j}^{\beta }(\theta
,B)=e^{iB}\,,  \label{Tb}
\end{eqnarray}
for this defect. The coupling constant is parameterized as in (\ref{coup}).
Evidently the ``unitarity'' (\ref{U2})-(\ref{U3}) and the crossing relations
(\ref{c1})-(\ref{c2}) are satisfied. Note that this is also an example for a
defect which breaks parity invariance, i.e., the left and right transmission
amplitudes are not identical. In the infinite lattice limit, i.e. when the
number of defects tends to infinity, we find 
\begin{equation}
\chi _{j/\bar{\jmath}}^{\beta }(\theta )=\cos (ky\mp B)\qquad \Rightarrow
\quad \lambda _{j/\bar{\jmath}}^{\beta }(\theta )\notin \Bbb{R},\Bbb{\,}%
\forall \,\theta ,B\,,
\end{equation}
which means that according to (\ref{unreal}) there are no forbidden energy
regimes.

\subsubsection{Energy insensitive defects, $\mathcal{D}^{\protect\gamma }(%
\bar{\protect\psi},\protect\psi )=g\bar{\protect\psi}\protect\gamma ^{5}%
\protect\psi $, $\mathcal{D}^{\protect\delta _{\pm }}(\bar{\protect\psi},%
\protect\psi )=g\bar{\protect\psi}(\protect\gamma ^{1}\pm \protect\gamma
^{5})\protect\psi $}

\vspace{-0.2cm} In comparison with the transparent defects the next
complication arises when the defect causes a phase shift independent of the
energy of the incoming particle. For $\mathcal{D}^{\gamma }(\bar{\psi},\psi
)=g\bar{\psi}\gamma ^{5}\psi $ we compute 
\begin{eqnarray}
R_{j}^{\gamma }(\theta ,B,-y)\!\!\! &=&\!\!\!\tilde{R}_{j}^{\gamma }(\theta
,-B,y)=R_{\bar{\jmath}}^{\gamma }(\theta ,B,y)=\tilde{R}_{\bar{\jmath}%
}^{\gamma }(\theta ,-B,-y)=ie^{2iym\sinh \theta }\tan B,\,\,\,\quad \\
T_{j}^{\gamma }(B)\!\!\! &=&\!\!\!T_{\bar{\jmath}}^{\gamma }(B)=\tilde{T}_{%
\bar{\jmath}}^{\gamma }(B)=\tilde{T}_{j}^{\gamma }(B)=\cos ^{-1}(B)\,.
\end{eqnarray}
In this case we observe that parity is broken for the reflection amplitudes,
i.e. $R\neq \tilde{R}$. The relations (\ref{U2})-(\ref{U3}) and (\ref{c1})-(%
\ref{c2}) for $y=0$ are satisfied. For $y=0$ none of the amplitudes depend
on the rapidities. In the infinite lattice limit we find 
\begin{equation}
\chi _{j}^{\gamma }(\theta )=\chi _{\bar{\jmath}}^{\gamma }(\theta )=\cos
ky\cos B<1\qquad \Bbb{\,}\forall \,\theta ,B\,,
\end{equation}
such that according to (\ref{bands}) there are no forbidden energy regimes.

For $\mathcal{D}^{\delta _{\pm }}(\bar{\psi},\psi )=g\bar{\psi}(\gamma
^{1}\pm \gamma ^{5})\psi $ we compute 
\begin{eqnarray}
R_{j}^{\delta _{\pm }}(\theta ,B,-y)\!\!\! &=&\!\!\!\tilde{R}_{j}^{\delta
_{\pm }}(\theta ,-B,y)\!=\!R_{\bar{\jmath}}^{\delta _{\pm }}(\theta
,B,y)\!=\!\tilde{R}_{\bar{\jmath}}^{\delta _{\pm }}(\theta ,-B,-y)\!=\!\!\pm
i\,\tan \frac{B}{2}e^{2iym\sinh \theta },\quad \,\,\,\,\, \\
T_{j}^{\delta _{\pm }}(B)\!\!\! &=&\!\!\!T_{\bar{\jmath}}^{\delta _{\pm
}}(-B)=\tilde{T}_{\bar{\jmath}}^{\delta _{\pm }}(B)=\tilde{T}_{j}^{\delta
_{\pm }}(-B)=1-2i\tan \frac{B}{2}\,.
\end{eqnarray}
These are examples in which parity is broken for the reflection as well as
for the transmission amplitudes. Again the relations (\ref{U2})-(\ref{U3})
and (\ref{c1})-(\ref{c2}) are satisfied when the defect is placed at the
origin and, as for $\mathcal{D}^{\gamma },$ when $y=0$ none of the
amplitudes depends on the rapidities. In this case we find in the infinite
lattice limit 
\begin{equation}
\chi _{j/\bar{\jmath}}^{\delta _{+}}(\theta )=\chi _{j/\bar{\jmath}}^{\delta
_{-}}(\theta )=\frac{\cos ky}{1\mp 2i\tan \frac{B}{2}}\qquad \Rightarrow
\quad \lambda _{j/\bar{\jmath}}^{\delta _{\pm }}(\theta )\notin \Bbb{R},\Bbb{%
\,}\forall \,\theta ,B\,,
\end{equation}
such that according to (\ref{bands}) there are no forbidden energy regions.

\subsubsection{Luttinger liquid type $\mathcal{D}^{\protect\varepsilon }(%
\bar{\protect\psi},\protect\psi )=\bar{\protect\psi}(g_{1}+g_{2}\protect%
\gamma ^{0})\protect\psi $}

\vspace{-0.2cm} \noindent When taking the conformal limit of a defect of the
type $\mathcal{D}^{\varepsilon }(\bar{\psi},\psi )=\bar{\psi}%
(g_{1}+g_{2}\gamma ^{0})\psi $ one obtains an impurity which played a role
in the context of Luttinger liquids \cite{Lutt} when setting the bosonic
number counting operator to zero, see e.g. \cite{Aff}. Besides $\mathcal{D}%
^{\alpha }(\bar{\psi},\psi )$ this is also an example of a defect for which
the potential is real. With (\ref{bcon}), (\ref{m1}) and (\ref{m2}) we
compute the related transmission and reflection amplitudes 
\begin{eqnarray}
R_{j}^{\varepsilon }(\theta ,g_{1},g_{2},-y)\!\!\! &=&\!\!\!\tilde{R}%
_{j}^{\varepsilon }(\theta ,g_{1},g_{2},y)=\frac{4i(g_{2}+g_{1}\cosh \theta
)e^{2iym\sinh \theta }}{(4+g_{1}^{2}-g_{2}^{2})\sinh \theta
-4i(g_{1}+g_{2}\cosh \theta )}\,, \\
R_{\bar{\jmath}}^{\varepsilon }(\theta ,g_{1},g_{2},-y)\!\!\! &=&\!\!\!%
\tilde{R}_{\bar{\jmath}}^{\varepsilon }(\theta ,g_{1},g_{2},y)=\frac{%
4i(g_{1}-g_{2}\cosh \theta )e^{-2iym\sinh \theta }}{(4+g_{1}^{2}-g_{2}^{2})%
\sinh \theta -4i(g_{1}-g_{2}\cosh \theta )}\,, \\
T_{j}^{\varepsilon }(\theta ,g_{1},g_{2})\!\!\! &=&\!\!\!\tilde{T}%
_{j}^{\varepsilon }(\theta ,g_{1},g_{2})=\frac{(4+g_{2}^{2}-g_{1}^{2})\sinh
\theta }{(4+g_{1}^{2}-g_{2}^{2})\sinh \theta -4i(g_{1}+g_{2}\cosh \theta )}%
\,\,, \\
T_{\bar{\jmath}}^{\varepsilon }(\theta ,g_{1},g_{2})\!\!\! &=&\!\!\!\tilde{T}%
_{\bar{\jmath}}^{\varepsilon }(\theta ,g_{1},g_{2})=\frac{%
(4+g_{2}^{2}-g_{1}^{2})\sinh \theta }{(4+g_{1}^{2}-g_{2}^{2})\sinh \theta
-4i(g_{1}-g_{2}\cosh \theta )}\,.
\end{eqnarray}
As we expect, since $\lim_{g_{2}\rightarrow 0}\mathcal{D}^{\varepsilon }(%
\bar{\psi},\psi )=\mathcal{D}^{\alpha }(\bar{\psi},\psi )$, we recover the
related results also for the $T/\tilde{T}$'s and $R/\tilde{R}$'s in (\ref
{ren})-(\ref{ten}). On the other hand, for $g_{1}$ $\rightarrow 0$ we obtain
the defect $\mathcal{D}^{\zeta }(\bar{\psi},\psi )=g_{2}\bar{\psi}\gamma
^{0}\psi $ for which the expressions simplify to 
\begin{eqnarray}
R_{j}^{\zeta }(\theta ,B,-y)\!\! &=&\!\!\tilde{R}_{j}^{\zeta }(\theta ,B,y)=%
\frac{-ie^{2iym\sinh \theta }\sin B}{\cos B\sinh \theta +i\sin B\cosh \theta 
}, \\
R_{\bar{\jmath}}^{\zeta }(\theta ,B,y)\!\! &=&\!\!\tilde{R}_{\bar{\jmath}%
}^{\zeta }(\theta ,B,-y)=\frac{ie^{2iym\sinh \theta }\sin B}{\cos B\sinh
\theta -i\sin B\cosh \theta }, \\
T_{j}^{\zeta }(\theta ,B)\!\! &=&\!\!\tilde{T}_{j}^{\zeta }(\theta ,B)=T_{%
\bar{\jmath}}^{\zeta }(\theta ,-B)=\tilde{T}_{\bar{\jmath}}^{\zeta }(\theta
,-B)=\frac{\sinh \theta }{\cos B\sinh \theta +i\sin B\cosh \theta }%
\,.\,\,\,\,\,\,
\end{eqnarray}
Where the effective coupling $B$ is given by (\ref{coup}) with $g\rightarrow
g_{2}$. The relations (\ref{U2})-(\ref{U3}) and (\ref{c1})-(\ref{c2}) may be
verified once again for $y=0$. In this case the infinite lattice limit leads
to forbidden energy regimes, since according to (\ref{bands}), the
rapidities have to respect the inequality 
\begin{equation}
(4+g_{1}^{2}-g_{2}^{2})\sinh \theta \cos ky+\sin ky(g_{1}\pm g_{2}\cosh
\theta )<(4+g_{2}^{2}-g_{1}^{2})\sinh \theta \quad \text{for }j,\bar{\jmath}%
\,,
\end{equation}
which possess non trivial solutions.

\subsubsection{The defects $\mathcal{D}^{\protect\eta _{\pm }}(\bar{\protect%
\psi},\protect\psi )=g\bar{\protect\psi}(1\pm \protect\gamma ^{5})/2\protect%
\psi $}

\vspace{-0.2cm} \noindent For this case we compute now 
\begin{eqnarray}
R_{j}^{\eta _{\pm }}(\theta ,B,y)\!\! &=&\!\!R_{\bar{\jmath}}^{\eta _{\pm
}}(\theta ,B,-y)=\frac{e^{\mp \theta }e^{-2iym\sinh \theta }}{i\cot
(B/2)\sinh \theta -1}\,, \\
\!\!\tilde{R}_{\bar{\jmath}}^{\eta _{\pm }}(\theta ,B,-y) &=&\tilde{R}%
_{j}^{\eta _{\pm }}(\theta ,B,y)\!\!=\frac{e^{\pm \theta }e^{2iym\sinh
\theta }}{i\cot (B/2)\sinh \theta -1}\,, \\
T_{j}^{\eta _{\pm }}(\theta ,B)\!\! &=&\!\!T_{\bar{\jmath}}^{\eta _{\pm
}}(\theta ,B)=\tilde{T}_{\bar{\jmath}}^{\eta _{\pm }}(\theta ,B)=\tilde{T}%
_{j}^{\eta _{\pm }}(\theta ,B)=\frac{1}{1\mp i\tan ^{-1}(B/2)\sinh
^{-1}(\theta )}\,,\,\,\,\,\,\,\,\,\,
\end{eqnarray}
which is once again in agreement with (\ref{U2})-(\ref{U3}) and (\ref{c1})-(%
\ref{c2}) for $y=0$. In the infinite lattice limit we obtain also in this
case forbidden energy regimes. The criterium (\ref{bands}) gives 
\begin{equation}
\pm \cos ky/2<\sinh \theta \tan B/2\sin ky/2\,,
\end{equation}
which has non-trivial solutions for the rapidities.

\subsubsection{The defects $\mathcal{D}^{\protect\lambda _{\pm }}(\bar{%
\protect\psi},\protect\psi )=g\bar{\protect\psi}(\protect\gamma ^{0}\pm 
\protect\gamma ^{1})/2\protect\psi $}

\vspace{-0.2cm} \noindent For this case we compute now 
\begin{eqnarray}
R_{j}^{\lambda _{\pm }}(\theta ,B,y)\!\!\! &=&\!\!\!R_{\bar{\jmath}%
}^{\lambda _{\pm }}(\theta ,B,-y)=\frac{e^{-2iym\sinh \theta }\tan \frac{B}{2%
}}{i\sinh \theta -\tan \frac{B}{2}\cosh \theta }\,, \\
\tilde{R}_{\bar{\jmath}}^{\lambda _{\pm }}(\theta ,B,y)\!\!\! &=&\!\!\!%
\tilde{R}_{j}^{\lambda _{\pm }}(\theta ,B,-y)=\frac{-e^{-2iym\sinh \theta
}\tan \frac{B}{2}}{i\sinh \theta +\tan \frac{B}{2}\cosh \theta }\,, \\
T_{j}^{\lambda _{\pm }}(\theta ,B)\!\!\! &=&\!\!\!T_{\bar{\jmath}}^{\lambda
_{\pm }}(\theta ,-B)=\tilde{T}_{\bar{\jmath}}^{\lambda _{\pm }}(-\theta ,B)=%
\tilde{T}_{j}^{\lambda _{\pm }}(-\theta ,-B)=\frac{(i\pm \tan \frac{B}{2}%
)\sinh \theta }{i\sinh \theta -\tan \frac{B}{2}\cosh \theta }%
.\,\,\,\,\,\,\,\,\,\,\,\,
\end{eqnarray}
The crossing-hermiticity and unitarity relations hold for $y=0$.

In principle we could of course prolong this list of defects and construct
their corresponding $R$'s and $T$'s. However, the main purpose of this
exercise was to review how the transmission and reflection amplitudes for a
defect may be obtained and also to give a brief account of some of their
characteristic features. Important to note is that indeed all variations of
possible parity breaking occur and one should keep therefore the discussion
generic in that sense. Note that when the defect is real, namely $\mathcal{D}%
^{\alpha }(\bar{\psi},\psi )$, $\mathcal{D}^{\varepsilon }(\bar{\psi},\psi
), $ parity invariance is preserved, which is a well known fact from quantum
mechanics (see e.g. \cite{Merz}). Complex potentials might look at first
sight somewhat unphysical from the energy spectrum point of view. However,
as is well-known for some bulk theories, such as for instance affine Toda
field theories with purely complex coupling constants, one can still
associate well defined quantum field theories to such Lagrangians and
construct even classically soliton solutions with real energies and momenta 
\cite{Tim}.

A classification scheme for possible defects which maintain integrability is
highly desirable. It is interesting to note that in the conformal limit, as
outlined before equation (\ref{LR0}), some of the defects, namely $\mathcal{D%
}^{\zeta }(\bar{\psi},\psi )$ and $\mathcal{D}^{\lambda _{\pm }}(\bar{\psi}%
,\psi ),$ become purely transmitting. Therefore, in contrast to first
impression, the folding idea \cite{WA} could be employed. We have now enough
examples at hand to use them in the following to determine the conductance
in a multiparticle system, which we shall do in two alternative ways.

\section{Conductance from the Landauer formula}

\vspace{-0.2cm} \setcounter{equation}{0}

\subsection{Conductance through an impurity}

\vspace{-0.2cm}

\noindent The most intuitive way to compute the conductance is via Landauer
transport theory \cite{Land}. Let us consider a set up as depicted in figure
3, that is we place a defect in the middle of a rigid bulk wire, where the
two halves might be at different temperatures.

\begin{center}
\includegraphics[width=9.0cm,height=5.7cm]{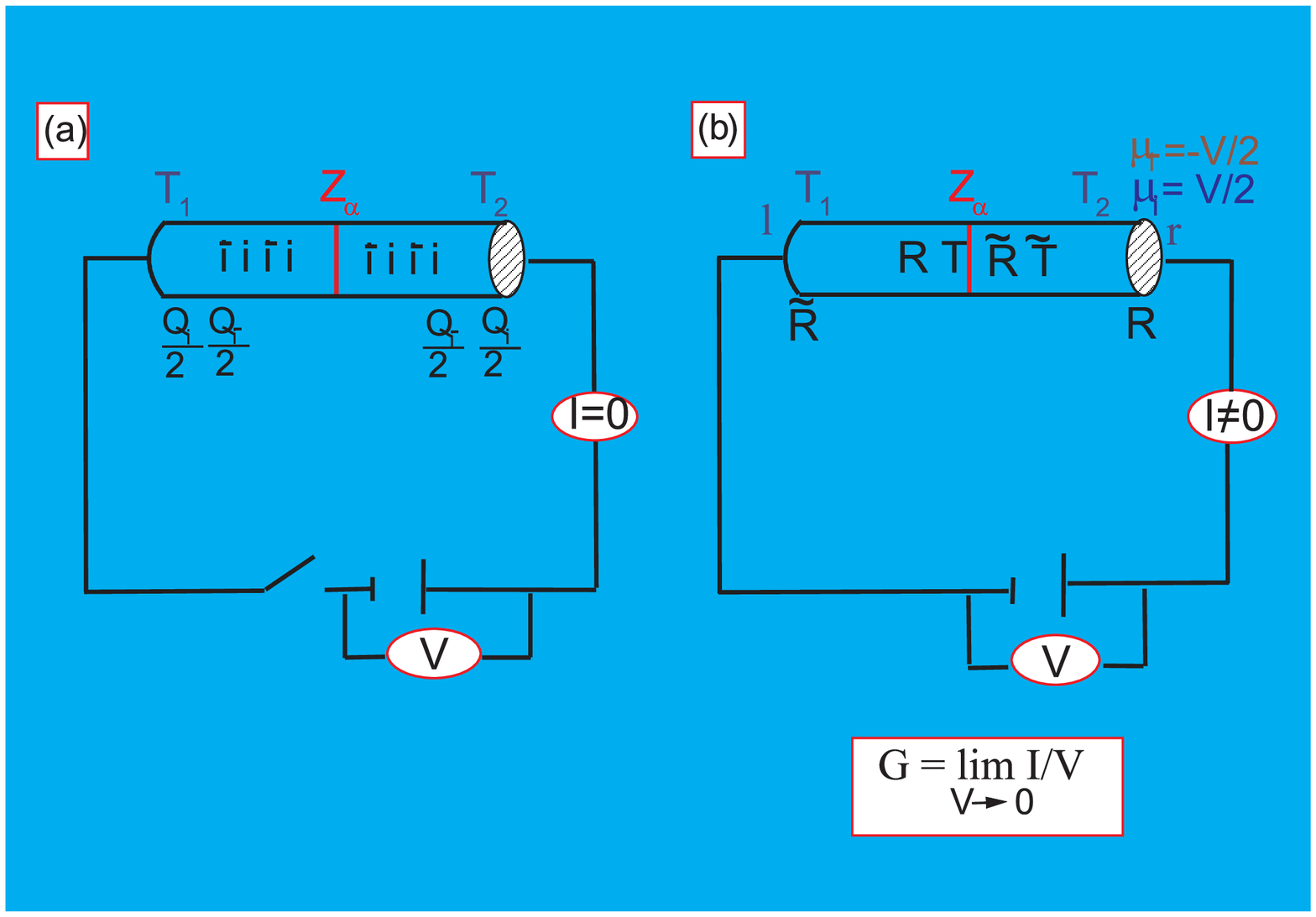}
\end{center}

\noindent {\small Figure 3: A conductance measurement. Part (a) represents
the initial condition with no current flowing, i.e., I=0 and part (b), $%
I\neq 0$. The defect is placed in the middle of the wire and the left and
right half are assumed to be at temperatures $T_{1}$ and $T_{2}$,
respectively. }

\medskip

\noindent The direct current $I$ through such a quantum wire can be computed
simply by determining the difference between the static charge distributions
at the right and left constriction of the wire, i.e. $I=Q_{r}-Q_{l}$. This
is based on the assumption \cite{FLS,LSS}, that $Q(t)\sim (Q_{r}-Q_{l})t\sim 
$ $(\rho _{r}-\rho _{l})t$, where the $\rho $s are the corresponding density
distribution functions. Placing an impurity in the middle of the wire, we
have to quantify the overall balance of particles of type $i$ and
anti-particles $\bar{\imath}$ carrying opposite charges $q_{i}=-q_{\bar{%
\imath}}\,$\ at the end of the wire at different potentials. This
information is of course encoded in the density distribution function $\rho
_{i}^{r}\left( \theta ,T,\mu _{i}\right) $.$\,$In the described set up half
of the particles of one type are already at the same potential at one of the
ends of the wire and the probability for them to reach the other is
determined by the transmission and reflection amplitudes through the
impurity. We assume that there is no effect coming from the constrictions of
the wire, i.e. they are purely transmitting surfaces with $T=\tilde{T}=1$.
One could, however, also consider a situation in which those constrictions
act as boundaries, namely purely reflecting surfaces. The situation could be
described with the same transport theory picture, see e.g. \cite
{FLS,FLS2,casa}, but then the conductance can only be non-vanishing if the
reflection amplitudes in the constrictions are non-diagonal in the particle
degrees of freedom, such as for instance for sine-Gordon \cite{GZ}, that is
in general affine Toda field theories with purely imaginary coupling
constant or in the massless limit folded purely reflecting (transmitting)
diagonal bulk theories.

According to the Landauer transport theory the direct current (DC) along the
wire is given by 
\begin{eqnarray}
I^{\mathbf{\alpha }} &=&\sum_{i}I_{i}^{\mathbf{\alpha }}(r,\mu _{i}^{l},\mu
_{i}^{r})=\sum_{i}\frac{q_{i}}{2}\int\limits_{-\infty }^{\infty }d\theta %
\left[ \rho _{i}^{r}(\theta ,r,\mu _{i}^{r})|T_{i}^{\mathbf{\alpha }}\left(
\theta \right) |^{2}-\rho _{i}^{r}(\theta ,r,\mu _{i}^{l})|\tilde{T}_{i}^{%
\mathbf{\alpha }}\left( \theta \right) |^{2}\right] ,\quad \,\,\,\,\,\,\,
\label{I11} \\
&=&I_{B}-\sum_{i}\frac{q_{i}}{2}\int\limits_{-\infty }^{\infty }d\theta %
\left[ \rho _{i}^{r}(\theta ,r,\mu _{i}^{r})|R_{i}^{\mathbf{\alpha }}\left(
\theta \right) |^{2}-\rho _{i}^{r}(\theta ,r,\mu _{i}^{l})|\tilde{R}_{i}^{%
\mathbf{\alpha }}\left( \theta \right) |^{2}\right] \,\,,  \label{I2}
\end{eqnarray}
where we assume here $T_{1}=T_{2}$. The relation (\ref{I2}) is obtained from
(\ref{I11}) simply by making use of the fact that $|R|^{2}+|T|^{2}=1$.
Equation (\ref{I2}) has the virtue that it extracts explicitly the bulk
contribution to the current which we refer to as $I_{B}$. There are some
obvious limits, namely a transparent and an impenetrable defect 
\begin{equation}
\lim_{|T^{\mathbf{\alpha }}|\rightarrow 1}I^{\mathbf{\alpha }}=I_{B}\qquad 
\text{and\qquad }\lim_{|T^{\mathbf{\alpha }}|\rightarrow 0}I^{\mathbf{\alpha 
}}=0\,,
\end{equation}
respectively. A short comment is needed on the validity of (\ref{I11}).
Apparently it suggests that when the parity between left and right
scattering is broken, there is the possibility of a net current even when an
external source is absent. In this picture we have of course not taken into
account that charged particles moving through the defect will alter the
potential, such that we did in fact not describe a perpetuum mobile. Thus
the limitation of our analysis is that $\mu _{i}^{l}-\mu _{i}^{r}$ has to be
much larger than the change in the potential induced by the moving particles.

Finally we want to compute the conductance from the DC current, which by
definition is obtained from 
\begin{equation}
G^{\mathbf{\alpha }}(r)=\sum\nolimits_{i}G_{i}^{\mathbf{\alpha }%
}(r)=\sum\nolimits_{i}\lim_{(\mu _{i}^{l}-\mu _{i}^{r})\rightarrow 0}I_{i}^{%
\mathbf{\alpha }}(r,\mu _{i}^{l},\mu _{i}^{r})\,/(\mu _{i}^{l}-\mu _{i}^{r})
\label{G}
\end{equation}
and is of course a property of the material itself and a function of the
temperature. In general the expressions in (\ref{I11}) tend to zero for
vanishing chemical potential difference such that the limit in (\ref{G}) is
non-trivial.

Thus from the knowledge of the transmission matrix and the density
distribution function we can compute the conductance. Having already
described how $T_{i}^{\mathbf{\alpha }}\left( \theta \right) $ can be
determined, we will now explain how $\rho _{i}^{r}(\theta ,r,\mu _{i})$ can
be evaluated by exploiting the integrability of the theory.

\subsection{Defect TBA equations}

\vspace{-0.2cm}

The thermodynamic Bethe ansatz is a powerful tool which may be used to
compute various thermodynamic properties of multi-particle systems which
interact via a factorizing scattering matrix \cite{TBAZam} and some chosen
statistics. In addition, it allows to check the theory for consistency and
to extract some distinct structural quantities such as the Virasoro central
in the ultraviolet limit. The original bulk formulation has been
accommodated to a situation which includes a purely transmitting defect \cite
{Marcio} and a boundary \cite{LMSS}. In this section we want to propose a
new formulation which is valid for a situation not treated before in this
context, namely when reflection and transmission occur simultaneously.

As usual we obtain the Bethe ansatz equation by dragging a particle along
the world line of length $L$. We introduce for convenience the following
shorthand notation for the product of various particle $Z_{i}(\theta )$ and
defect operators $Z_{\alpha }$ 
\begin{equation}
Z_{k_{1},\alpha _{1};k_{2},\alpha _{2}\ldots k_{n},\alpha _{n}}^{\mu
_{1}\ldots \mu _{N}}:=Z_{\mu _{1}}(\theta _{\mu _{1}})\ldots Z_{\mu
_{k_{1}}}(\theta _{\mu _{k_{1}}})Z_{\alpha _{1}}\ldots Z_{\mu
_{k_{n}}}(\theta _{\mu _{k_{n}}})Z_{\alpha _{n}}\ldots Z_{\mu _{N}}(\theta
_{\mu _{N}}).
\end{equation}
Then we compute the braiding of a particle operator of type $i$ and the
product of $N$ further particles $Z_{\mu _{1}}\ldots Z_{\mu _{N}}$ with one
defect $Z_{\alpha }$ situated on the right of the particle $Z_{\mu _{k}}$ by
using the ZF-algebra (\ref{Z1}) and (\ref{Z2}) 
\begin{eqnarray}
Z_{i}(\theta _{i})Z_{k,\alpha }^{\mu _{1}\ldots \mu _{N}} &=&Z_{k,\alpha
}^{\mu _{1}\ldots \mu _{N}}Z_{i}(\theta _{i})\tilde{F}_{i\alpha
}-Z_{k,\alpha }^{\mu _{1}\ldots \mu _{N}}Z_{i}(-\theta _{i})\tilde{G}%
_{i\alpha }\,,  \label{p1} \\
Z_{k,\alpha }^{\mu _{1}\ldots \mu _{N}}Z_{i}(\theta _{i}) &=&Z_{i}(\theta
_{i})Z_{k,\alpha }^{\mu _{1}\ldots \mu _{N}}F_{i\alpha }-Z_{i}(-\theta
_{i})Z_{k,\alpha }^{\mu _{1}\ldots \mu _{N}}G_{i\alpha }\,.  \label{p2}
\end{eqnarray}
We abbreviated here 
\begin{eqnarray}
\tilde{F}_{i}^{\alpha }\!\!\! &=&\!\!\!\frac{1}{\tilde{T}_{i}^{\alpha
}(-\theta _{i})}\prod_{l=1}^{N}S_{i\mu _{l}}(\theta _{i\mu _{l}})\,,\text{%
\quad }\tilde{G}_{i}^{\alpha }=\frac{\tilde{R}_{i}^{\alpha }(-\theta _{i})}{%
\tilde{T}_{i}^{\alpha }(-\theta _{i})}\prod_{l=1}^{k}S_{i\mu _{l}}(\theta
_{i\mu _{l}})\prod_{l=k+1}^{N}S_{i\mu _{l}}(-\hat{\theta}_{i\mu
_{l}})\,,\,\,\,\,\, \\
F_{i}^{\alpha }\!\!\! &=&\!\!\!\frac{1}{T_{i}^{\alpha }(\theta _{i})}%
\prod_{l=1}^{N}S_{\mu _{l}i}(\theta _{\mu _{l}i})\,,\text{\quad \thinspace
\thinspace \thinspace \thinspace \thinspace }G_{i}^{\alpha }=\frac{%
R_{i}^{\alpha }(\theta _{i})}{T_{i}^{\alpha }(\theta _{i})}%
\prod_{l=1}^{k}S_{\mu _{l}i}(\hat{\theta}_{\mu
_{l}i})\prod_{l=k+1}^{N}S_{\mu _{l}i}(\theta _{\mu _{l}i})\,\,.
\end{eqnarray}
Being on a circle of length $L$, we can make the usual assumption on the
Bethe wavefunction, see e.g. \cite{TBAZam}, which is captured in the
requirement 
\begin{equation}
Z_{i}(\theta )Z_{k,\alpha }^{\mu _{1}\ldots \mu _{N}}=Z_{k,\alpha }^{\mu
_{1}\ldots \mu _{N}}Z_{i}(\theta )\,\exp (-iLm_{i}\sinh \theta )\,.
\end{equation}
Using this monodromy property together with the braiding relations (\ref{p1}%
), (\ref{p2}) and the unitarity relation (\ref{U2}), we obtain 
\begin{equation}
\prod_{l=1}^{N}\frac{S_{li}(\hat{\theta}_{li})}{S_{li}(\theta _{li})}\left(
\prod_{l=1}^{N}S_{li}(\theta _{li})-\frac{e^{iLm_{i}\sinh \theta _{i}}}{%
\tilde{T}_{i}^{\alpha }(-\theta _{i})}\right) =\frac{T_{i}^{\alpha }(-\theta
_{i})}{\tilde{T}_{i}^{\alpha }(-\theta _{i})}\left( \frac{e^{-iLm_{i}\sinh
\theta _{i}}}{T_{i}^{\alpha }(\theta _{i})}-\prod_{l=1}^{N}S_{il}(\theta
_{il})\right) \,.  \label{BA}
\end{equation}
Viewing the subscripts as entire spaces rather than components, equation (%
\ref{BA}) corresponds to the Bethe ansatz equation with simultaneously
occurring transmission and reflection amplitudes for the generic, that is
also the non-diagonal, case. We restrict it here to the diagonal case and
can therefore use the constraints (\ref{S}), such that the bulk scattering
matrix becomes rapidity independent and the relation (\ref{BA}) may be
re-written as 
\begin{equation}
1=e^{iLm_{i}\sinh \theta }D_{i\alpha }^{\pm }(\theta
)\prod\nolimits_{l=1}^{N}S_{il}  \label{Ban}
\end{equation}
where 
\begin{equation}
D_{i\alpha }^{\pm }(\theta )=\frac{\tilde{T}_{i}^{\alpha }(\theta
)+T_{i}^{\alpha }(\theta )\prod\nolimits_{l=1}^{N}S_{il}^{2}}{2}\pm \frac{1}{%
2}\left[ \left( \tilde{T}_{i}^{\alpha }(\theta )+T_{i}^{\alpha }(\theta
)\prod\limits_{l=1}^{N}S_{il}^{2}\right) ^{2}-\frac{4T_{i}^{\alpha }(\theta
)\prod\nolimits_{l=1}^{N}S_{il}^{2}}{T_{i}^{\alpha }(-\theta )}\right] ^{%
\frac{1}{2}}.  \label{ds}
\end{equation}
For consistency reasons it is instructive to consider the limit when the
reflection amplitude tends to zero. In that case we can employ the relations
(\ref{U1})-(\ref{U3}) and may take the square root in (\ref{ds}), such that
we obtain from (\ref{Ban}) the two equations 
\begin{equation}
R,\tilde{R}\rightarrow 0:\qquad 1=e^{iLm_{i}\sinh \theta }\tilde{T}%
_{i}^{\alpha }(\theta )\prod_{l=1}^{N}S_{il}\,,\text{\quad }%
1=e^{-iLm_{i}\sinh \theta }T_{i}^{\alpha }(\theta )\prod_{l=1}^{N}S_{li}\,\,.
\label{lba}
\end{equation}
This means we recover the Bethe ansatz equations for a purely transmitting
defect, which were originally proposed by Martins in \cite{Marcio}. The two
signs in (\ref{ds}) capture the breaking of parity invariance in the
limiting case, i.e. the two equations in (\ref{lba}) correspond to taking
the particle either clockwise or anti-clockwise around the world line as
formulated for the parity breaking case for the first time in \cite{CFKM}.
We do not expect to recover from here the equations for a purely reflecting
boundary which were suggested in \cite{LMSS}, since the equations (\ref{p1})
and (\ref{p2}) do not make sense in the limit $T,\tilde{T}\rightarrow 0$.
For $\prod\nolimits_{l=1}^{N}S_{il}^{2}=1,$ i.e. the free Boson and Fermion,
we can exploit the fact that (\ref{Ban}) with (\ref{ds}) look formally
precisely like the Bethe ansatz equations for a purely transmitting defect.
If we want to exploit this analogy we should of course be concerned about
the question whether $D_{j\alpha }^{\pm }(\theta )$ is a meromorphic
function. Assuming parity invariance, we may take the square root 
\begin{equation}
D_{j\alpha }^{\pm }(\theta )=T_{j}^{\alpha }(\theta )\,\pm R_{j}^{\alpha
}(\theta )\,\quad \quad \quad \text{for\quad }T=\tilde{T},R=\tilde{R}\,.
\label{s}
\end{equation}
The matrix $D_{j\alpha }^{\pm }(\theta )$ has now the usual properties,
namely it is unitarity in the sense that $D_{j\alpha }^{\pm }(\theta
)D_{j\alpha }^{\pm }(-\theta )=1$. It follows further from (\ref{s}), (\ref
{c1}) and (\ref{c2}) that the hermiticity relation $D_{j\alpha }^{\pm
}(\theta )=D_{j\alpha }^{\pm }(-\theta )^{\ast }$ and the crossing relations 
$D_{\bar{\jmath}\alpha }^{\pm }(\theta )=D_{j\alpha }^{\mp }(i\pi -\theta )$
and $D_{\bar{\jmath}\alpha }^{\pm }(\theta )=D_{j\alpha }^{\pm }(i\pi
-\theta )$ hold for the free Fermion and Bosons, respectively.

Let us now carry out the thermodynamic limit in the usual way, namely by
increasing the particle number and the system size in such a way that their
mutual ratio remains finite. The amount of defects is kept constant in this
limit, such that there is no contribution to the TBA-equations from the
defect in that situation, see also \cite{Marcio} where the same argument was
employed. Hence this means that essentially we can employ the usual bulk TBA
analysis when the considerations are carried out not per unit length.

Let us therefore recall the main equations of the TBA analysis. For more
details on the derivation see \cite{TBAZam} and in particular for the
introduction of the chemical potential see \cite{TBAKM}. The main input into
the entire analysis is the dynamical interaction, which enters via the
logarithmic derivative of the scattering matrix $\varphi _{ij}(\theta
)=-id\ln S_{ij}(\theta )/d\theta $ and the assumption on the statistical
interaction, which we take to be fermionic. As usual \cite{TBAZam,TBAKM}, we
take the logarithmic derivative of the Bethe ansatz equation (\ref{Ban}) and
relate the density of states $\rho _{i}(\theta ,r)$ for particles of type $i$
as a function of the inverse temperature $r=1/T$ to the density of occupied
states $\rho _{i}^{r}(\theta ,r)$ 
\begin{equation}
\rho _{i}(\theta ,r)=\frac{m_{i}}{2\pi }\cosh \theta
+\sum\nolimits_{j}[\varphi _{ij}\ast \rho _{i}^{r}](\theta )\,.  \label{rho}
\end{equation}
By $\left( f\ast g\right) (\theta )$ $:=1/(2\pi )\int d\theta ^{\prime
}f(\theta -\theta ^{\prime })g(\theta ^{\prime })$ we denote as usual the
convolution of two functions. The mutual ratio of the densities serves as
the definition of the so-called pseudo-energies $\varepsilon _{i}(\theta ,r)$%
\begin{equation}
\frac{\rho _{i}^{r}(\theta ,r)}{\rho _{i}(\theta ,r)}=\frac{e^{-\varepsilon
_{i}(\theta ,r)}}{1+e^{-\varepsilon _{i}(\theta ,r)}}\,,  \label{dens}
\end{equation}
which have to be positive and real. At thermodynamic equilibrium one obtains
then the TBA-equations, which read in these variables and in the presence of
a chemical potential $\mu _{i}$%
\begin{equation}
rm_{i}\cosh \theta =\varepsilon _{i}(\theta ,r,\mu _{i})+r\mu
_{i}+\sum\nolimits_{j}[\varphi _{ij}\ast \ln (1+e^{-\varepsilon
_{j}})](\theta )\,,  \label{TBA}
\end{equation}
where $r=m/T$, $m_{l}\rightarrow m_{l}/m$, $\mu _{i}\rightarrow \mu _{i}/m$,
with $m$ being the mass of the lightest particle in the model. It is
important to note that $\mu _{i}$ is restricted to be smaller than 1. This
follows immediately from (\ref{TBA}) by recalling that $\varepsilon _{i}\geq
0$ and that for $r$ large $\varepsilon _{i}(\theta ,r,\mu _{i})$ tends to
infinity. As pointed out already in \cite{TBAZam} (here just with the small
modification of a chemical potential), the comparison between (\ref{TBA})
and (\ref{rho}) leads to the useful relation 
\begin{equation}
\rho _{i}(\theta ,r,\mu _{i})=\frac{1}{2\pi }\left( \frac{d\varepsilon
_{i}(\theta ,r,\mu _{i})}{dr}+\mu _{i}\right) \,.  \label{rhoe}
\end{equation}
The main task is therefore first to solve (\ref{TBA}) for the
pseudo-energies from which then all densities can be reconstructed.

\subsection{Thermodynamic quantities\ }

\vspace{-0.2cm}

\noindent Treating the equations (\ref{Ban}) and (\ref{ds}) in the mentioned
analogy we can also construct various thermodynamic quantities. It should be
stressed that these quantities are computed per unit length. Similarly as
the expression found in \cite{Marcio} for a purely transmitting defect the
free energy is 
\begin{equation}
F(r)=-\frac{1}{\pi r}\sum_{l,\alpha }\hat{m}_{l}\int\nolimits_{0}^{\infty
}d\theta \,[\cosh \theta +m^{-1}\varphi _{l\alpha }(\theta )]\,\ln [1+\exp
(-rm\cosh \theta )]\,.  \label{scale}
\end{equation}
It is made up of two parts, one coming from the bulk and one including the
data of the defect in form of $\varphi _{l\alpha }(\theta )=-id\ln
D_{l\alpha }(\theta )/d\theta $. From equation (\ref{scale}) we also see
that when taking the mass scale to be large in comparison to the dominating
scale in the defect, the latter contribution to the scaling function becomes
negligible with regard to the bulk and vice versa.

\subsection{The high temperature regime}

\vspace{-0.2cm} \noindent Since the physical quantities require a solution
of the TBA-equations, which up to now, due to their non-linear nature, can
only be solved numerically, we have to resort in general to a numerical
analysis to obtain the conductance for some concrete theories. However,
there exist various approximations for different special situations, such as
the high temperature regime. For large rapidities and small $r$, it is known 
\cite{TBAZam} (here we only need the small modification of the introduction
of a chemical potential $\mu _{i}$) that the density of states can be
approximated by 
\begin{equation}
\rho _{i}(\theta ,r,\mu _{i})\sim \frac{m_{i}}{4\pi }e^{|\theta |}\sim \frac{%
1}{2\pi r}\epsilon (\theta )\frac{d\varepsilon _{i}(\theta ,r,\mu _{i})}{%
d\theta }\,,  \label{rr}
\end{equation}
where $\epsilon (\theta )=\Theta (\theta )-\Theta (-\theta )$ is the step
function, i.e. $\epsilon (\theta )=1$ for $\theta >0$ and $\epsilon (\theta
)=-1$ for $\theta <0$. In equation (\ref{dens}), we assume that in the large
rapidity regime $\rho _{i}^{r}(\theta ,r,\mu _{i})$ is dominated by (\ref{rr}%
) and in the small rapidity regime by the Fermi distribution function.
Therefore 
\begin{equation}
\rho _{i}^{r}(\theta ,r,\mu _{i})\sim \frac{1}{2\pi r}\epsilon (\theta )%
\frac{d}{d\theta }\ln \left[ 1+\exp (-\varepsilon _{i}(\theta ,r,\mu _{i}))%
\right] \,\,.
\end{equation}
Using this expression in equation (\ref{I11}), we approximate the direct
current in the ultraviolet by 
\begin{equation}
\lim\limits_{r\rightarrow 0}I_{i}^{\mathbf{\alpha }}(r,\mu _{i})\sim \frac{%
q_{i}}{4\pi r}\int\limits_{-\infty }^{\infty }d\theta \ln \left[ \frac{%
1+\exp (-\varepsilon _{i}(\theta ,r,\mu _{i}^{l}))}{1+\exp (-\varepsilon
_{i}(\theta ,r,\mu _{i}^{r}))}\right] \,\frac{d\left[ \epsilon (\theta
)\,|T_{i}^{\mathbf{\alpha }}(\theta )|^{2}\right] }{d\theta }\,,  \label{ya}
\end{equation}
after a partial integration. For simplicity we also assumed here parity
invariance, that is $|T_{i}^{\mathbf{\alpha }}(\theta )|=|\tilde{T}_{i}^{%
\mathbf{\alpha }}(\theta )|$. The derivation of the analogue to (\ref{ya})
for the situation when parity is broken is of course similar. Taking now the
potentials at the end of the wire to be $\mu _{i}^{r}=-\mu _{i}^{l}=V/2$,
the conductance reads in this approximation 
\begin{equation}
\lim\limits_{r\rightarrow 0}G_{i}^{\mathbf{\alpha }}(r)\sim \frac{q_{i}}{%
2\pi r}\int\limits_{-\infty }^{\infty }d\theta \frac{1}{1+\exp [\varepsilon
_{i}(\theta ,r,0)]}\left. \frac{d\varepsilon _{i}(\theta ,r,V/2)}{dV}\right|
_{V=0}\frac{d\left[ \epsilon (\theta )\,|T_{i}^{\mathbf{\alpha }}(\theta
)|^{2}\right] }{d\theta }\,.
\end{equation}
In order to evaluate these expressions further, we need to know explicitly
the precise form of the transmission matrix, i.e. the concrete form of the
defect. An interesting situation occurs when the defect is transparent or
rapidity independent, that is $|T_{i}^{\mathbf{\alpha }}(\theta
)|\rightarrow |T_{i}^{\mathbf{\alpha }}|$, in which case we can pursue the
analysis further. Noting that $d\epsilon (\theta )/d\theta =2\delta (\theta
) $, we obtain 
\begin{equation}
\lim\limits_{r\rightarrow 0}G_{i}^{\mathbf{\alpha }}(r)\sim \frac{q_{i}}{\pi
r}\frac{|T_{i}^{\mathbf{\alpha }}|^{2}}{1+\exp \varepsilon _{i}(0,r,0)}%
\left. \frac{d\varepsilon _{i}(0,r,V/2)}{dV}\right| _{V=0}\,.  \label{g0}
\end{equation}
The derivative $d\varepsilon _{i}(0,r,V/2)/dV$ can be obtained by solving
recursively 
\begin{equation}
\frac{d\varepsilon _{i}(0,r,V/2)}{dV}=-\frac{r}{2}-\sum_{j}N_{ij}\frac{1}{%
1+\exp \varepsilon _{j}(0,r,V/2)]}\frac{d\varepsilon _{j}(0,r,V/2)}{dV}\,,
\end{equation}
which results form a computation similar to a standard one in this context 
\cite{TBAZam} leading to the so-called constant TBA-equations. Here only the
asymptotic phases of the scattering matrix enter via $N_{ij}=\lim_{\theta
\rightarrow \infty }[\ln [S_{ij}(-\theta )/S_{ij}(\theta )]]/2\pi i$. The
values of $\varepsilon _{i}(0,r,0)$ needed in (\ref{g0}) can be obtained for
small $r$ in the usual way from the standard constant TBA-equations.

\subsection{Free Fermion with defects\ \ }

\vspace{-0.2cm}

\noindent Let us exemplify the general formulae once more with the free
Fermion. First of all we note that in this case in the TBA-equations (\ref
{TBA}) the kernel $\varphi _{ij}$ is vanishing and the equation is simply
solved by 
\begin{equation}
\varepsilon _{i}(\theta ,r,\mu _{i})=rm_{i}\cosh \theta -r\mu _{i}\,.
\label{ffe}
\end{equation}
Therefore, we have explicit functions for the densities with (\ref{rhoe})
and (\ref{dens}) 
\begin{equation}
\rho _{i}\left( \theta ,r,\mu _{i}\right) =\frac{1}{2\pi }m_{i}\cosh \theta
\qquad \text{and\qquad }\rho _{i}^{r}\left( \theta ,r,\mu _{i}\right) =\frac{%
m_{i}\cosh \theta /2\pi }{1+\exp (rm_{i}\cosh \theta -r\mu _{i})}\,.
\label{rfe}
\end{equation}
According to (\ref{I11}) the direct current reads 
\begin{eqnarray}
I^{\mathbf{\alpha }}(r,V) &=&\frac{q_{i}}{2}\int\limits_{-\infty }^{\infty
}d\theta \left[ \rho _{\bar{\imath}}^{r}\left( \theta ,r,V/2\right) |T_{\bar{%
\imath}}^{\mathbf{\alpha }}\left( \theta \right) |^{2}-\rho _{i}^{r}\left(
\theta ,r,-V/2\right) |T_{i}^{\mathbf{\alpha }}\left( \theta \right)
|^{2}\right.  \nonumber \\
&&\left. \quad \qquad \quad -\rho _{\bar{\imath}}^{r}\left( \theta
,r,-V/2\right) |\tilde{T}_{\bar{\imath}}^{\mathbf{\alpha }}\left( \theta
\right) |^{2}+\rho _{i}^{r}\left( \theta ,r,V/2\right) |\tilde{T}_{i}^{%
\mathbf{\alpha }}\left( \theta \right) |^{2}\right] \,\,.
\end{eqnarray}
Using atomic units $m_{e}=e=%
h\hskip-.2em\llap{\protect\rule[1.1ex]{.325em}{.1ex}}\hskip.2em%
=m_{i}=q_{i}=1$, we obtain explicitly with (\ref{rfe}) 
\begin{equation}
I^{\mathbf{\alpha }}(r,V)=\frac{1}{\pi }\int\limits_{0}^{\infty }d\theta 
\frac{\cosh \theta \sinh (rV/2)\,|T^{\mathbf{\alpha }}\left( \theta \right)
|^{2}}{\cosh (r\cosh \theta )+\cosh (rV/2)},  \label{44}
\end{equation}
for $|T_{\bar{\imath}}^{\mathbf{\alpha }}\left( \theta \right) |=|T_{i}^{%
\mathbf{\alpha }}\left( \theta \right) |=|\tilde{T}_{\bar{\imath}}^{\mathbf{%
\alpha }}\left( \theta \right) |=|\tilde{T}_{i}^{\mathbf{\alpha }}\left(
\theta \right) |=|T^{\mathbf{\alpha }}\left( \theta \right) |\,$. Then by (%
\ref{G}) the conductance results to 
\begin{equation}
G^{\mathbf{\alpha }}(r)=rm\frac{e^{2}}{h}\int\limits_{0}^{\infty }d\theta 
\frac{\cosh \theta \,\left| T^{\mathbf{\alpha }}\left( \theta \right)
\right| ^{2}}{1+\cosh (rm\cosh \theta )}\,  \label{gg}
\end{equation}
in this case. We have re-introduced dimensional quantities instead of atomic
units to be able to match with some standard results from the literature.
The most characteristic features can actually be captured when we carry out
the massless limit as indicated in section 2.3.2, which can be done even
analytically. Substituting $t=e^{\theta }$, we obtain 
\begin{equation}
\lim_{m\rightarrow 0}G^{\mathbf{\alpha }}(r)\sim \frac{e^{2}}{h}%
\int\limits_{0}^{\infty }dt\frac{\,|T_{L/R}^{\mathbf{\alpha }}(t\,y/r)|^{2}}{%
1+\cosh (t)}=\frac{e^{2}}{h}\left\{ 
\begin{array}{l}
\overline{|T_{L/R}^{\mathbf{\alpha }}(t\,y/r)|^{2}}\qquad \quad \,\,\,\,\,%
\text{for }y\gg r \\ 
|T_{L/R}^{\mathbf{\alpha }}(y/r=0)|^{2}\quad \quad \,\,\,\text{for }y\ll r
\end{array}
\right. \,.  \label{massl}
\end{equation}
We have identified here two distinct regions. When $y\ll r$ we can replace
the left/right transmission amplitudes by their values at $y/r=0$. When $%
y\gg r$ the transmission amplitudes enter the expression as a strongly
oscillatory function in which $y/r$ plays the role of the frequency. It is
then a good approximation to replace this function by its means value as
indicated by the overbar. It is straightforward to extend the expression (%
\ref{massl}) to the case when the assumption on $T^{\mathbf{\alpha }}$ in (%
\ref{44}) is relaxed and to the case with different values of $y$. To
proceed further we need to specify the defect.

\subsubsection{Energy insensitive defects, $\mathcal{D}^{0}(\bar{\protect\psi%
},\protect\psi )=0$, $\mathcal{D}^{\protect\beta }(\bar{\protect\psi},%
\protect\psi )$, $\mathcal{D}^{\protect\gamma }(\bar{\protect\psi},\protect%
\psi )$, $\mathcal{D}^{\protect\delta _{\pm }}(\bar{\protect\psi},\protect%
\psi )$}

\vspace{-0.2cm} Let us first consider the easiest example, which supports
the general working of the method. When the defect is transparent, i.e., $%
|T^{\mathbf{\alpha }}|=1$, we can compute the expression for the conductance
(\ref{gg}) directly in the large temperature limit and obtain the well known
behaviour \cite{KF} 
\begin{equation}
\lim_{r\rightarrow 0,|T^{\mathbf{\alpha }}|\rightarrow 1}G^{\mathbf{\alpha }%
}(r)\sim \frac{e^{2}}{h}(1-\frac{rm}{2})\,.  \label{hh}
\end{equation}
Alternatively we obtain the expression (\ref{hh}) also from equation (\ref
{g0}) and (\ref{ffe}). In the massless limit of (\ref{massl}) we obtain $%
e^{2}/h$ which coincides with the result in \cite{FLS}. However, we should
stress that we consider here purely massive cases and the massless limit
only serves as a benchmark. Note that a transparent defect in this context
does not necessarily mean the absence of the defect. Considering for
instance the defect $\mathcal{D}^{\beta }(\bar{\psi},\psi )$, we compute
with (\ref{Rb}) and (\ref{Tb}) the same conductance as if there was no
defect at all. Similarly simple are the computations for the defects $%
\mathcal{D}^{\gamma }(\bar{\psi},\psi )$, $\mathcal{D}^{\delta _{\pm }}(\bar{%
\psi},\psi )$. We simply get 
\begin{equation}
G^{0}(r)=G^{\beta }(r)=G^{\gamma }(r)\cos ^{2}B=G^{\delta _{\pm
}}(r)/(1+4\tan ^{2}B/2)\,\,=\frac{e^{2}}{h}.  \label{moorhuhn}
\end{equation}
Since the amplitudes do not depend on the rapidities, the TBA-kernel is zero
and there is no contribution from this defect to the free energy, even unit
length.

\subsubsection{The energy operator defect $\mathcal{D}^{\protect\alpha }(%
\bar{\protect\psi},\protect\psi )=g\bar{\protect\psi}\protect\psi $}

\vspace{-0.2cm}

\noindent For this defect the computation of the conductance according to (%
\ref{gg}) is more involved. The results of our numerical analysis of the
expression (\ref{gg}) are depicted in figure 4.

\begin{center}
\includegraphics[width=15.cm,height=5.7cm]{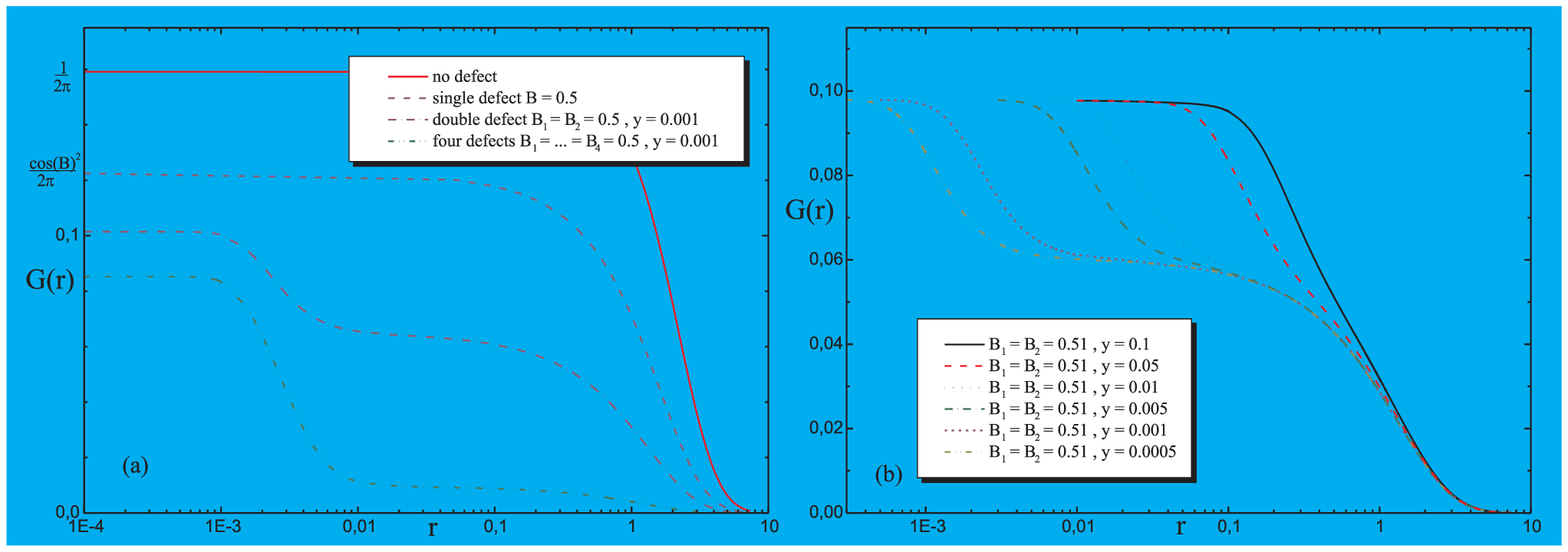}
\end{center}

\noindent {\small Figure 4: Conductance }$G(r)${\small \ for the complex
free Fermion with the energy operator defects as a function of the inverse
temperature }$r,$ {\small for fixed effective coupling constant }$B${\small %
\ and (a) for varying amounts of defects $\ell =0,1,2,4$. (b) for $\ell =2$
for varying distances }$y${\small .}

\medskip

We observe several distinct features. First of all it is naturally to be
expected that when we increase the number of defects the resistance will
grow. This is confirmed, as for fixed temperature and increasing number of
defects, the conductance decreases. Second we see several well extended
plateaux. They can be reproduced with the analytical expressions obtained in
the massless limit (\ref{massl}). To be able to compare with (\ref{gg}) we
re-introduce atomic units for convenience, i.e. $e^{2}/h\rightarrow 1/2\pi $%
. For a single defect there is only one plateaux and from (\ref{massl}) we
obtain with (\ref{LR0}) 
\begin{equation}
G^{\alpha }(r)\sim \frac{\cos ^{2}B}{2\pi }\,.  \label{g1}
\end{equation}
For $B=0.5$ the value $0.1226$ is well reproduced in figure 4(a). The lower
lying plateaux correspond to the region when $y\ll r$. In that case we
obtain from (\ref{massl}) together with the expressions (\ref{LR2})-(\ref
{LR5}) for a double and four defects 
\begin{eqnarray}
G^{\alpha _{1}\alpha _{2}}(r) &\sim &\frac{1}{2\pi }\left( \frac{\cos ^{2}B}{%
1+\sin ^{2}B}\right) ^{2}\qquad \quad \,\quad \quad \qquad \quad \qquad
\,\,\,\text{for }y\ll r\,,  \label{o1} \\
G^{\alpha _{1}\alpha _{2}\alpha _{3}\alpha _{4}}(r) &\sim &\frac{1}{2\pi }%
\left( \frac{\cos ^{4}B}{\cos ^{4}B-2(1+\sin ^{2}B)^{2}}\right) ^{2}\,\quad
\quad \qquad \,\,\,\,\text{for }y\ll r.  \label{o2}
\end{eqnarray}
For $B=0.5$ the values $0.0624$ and $0.0095$ are well reproduced in figure
4(a) for $\ell =2$ and $\ell =4$, respectively. The plateaux extending to
the ultraviolet regime result from (\ref{massl}) and by taking in (\ref{LR2}%
)-(\ref{LR5}) the mean values 
\begin{eqnarray}
G^{\alpha _{1}\alpha _{2}}(r) &\sim &\frac{2}{\pi }\frac{1+\sin ^{4}B}{(\cos
^{2}(2B)-3)^{2}}\,,\qquad \quad \,\quad \qquad \quad \,\,\,\,\,\,\,\,\quad 
\text{for }y\gg r\,,  \label{o3} \\
G^{\alpha _{1}\alpha _{2}\alpha _{3}\alpha _{4}}(r) &\sim &\frac{1}{4\pi }+%
\frac{\cos ^{8}B}{4\pi \lbrack \cos ^{4}B-2(1+\sin ^{2}B)^{2}]^{2}}%
,\,\,\quad \,\,\quad \text{for }y\gg r.  \label{o4}
\end{eqnarray}
Also in this case the values for $B=0.5$, i.e., $0.110784$ and $0.084311$
for $\ell =2$ and $\ell =4$, respectively, match very well with the
numerical analysis. Finally we have to explain the reason for the increase
from one to the next plateaux and why the curves are shifted precisely in
the way as indicated in figure 4(b) when we change the distance between the
defects. This phenomenon is attributed to resonances as we shall discuss in
more detail in the next subsection.

\subsubsection{Resonances versus unstable particles}

\vspace{-0.2cm}

In \cite{CF12} we demonstrated that resonances may be described similarly as
unstable particles. The latter provide an intuitively very clear picture
which explains the relatively sharp onset of the conductance with increasing
temperature. The temperature at which this onset occurs, say $T_{C}$ can be
attributed directly to the energy scale at which the unstable particle is
formed, since then it starts to participate in the conducting process. The
Breit-Wigner formula \cite{BW} constitutes in this case a relation for the
mass $M$ and the decay width $\Gamma $ of an unstable particle. Supposing
that in the scattering process between particles of type $i$ and $j$ an
unstable particle can be formed, this is reflected by a pole in $%
S_{ij}(\theta )$ at $\theta =\sigma -i\bar{\sigma}$. Then, for large values
of the resonance parameter $\sigma $ one may approximate 
\begin{equation}
M^{2}\approx 1/2m_{i}m_{j}(1+\cos \bar{\sigma})\exp |\sigma |\qquad \text{%
and\qquad }\Gamma ^{2}\approx 2m_{i}m_{j}(1-\cos \bar{\sigma})\exp |\sigma
|\,.  \label{BW}
\end{equation}
Since a renormalization group flow is provided by $M\rightarrow r\,M$, one
should observe that the quantities $M\sim r_{1}e^{\sigma
_{1}/2}=r_{2}e^{\sigma _{2}/2}$ and $\Gamma \sim r_{1}e^{\sigma
_{1}/2}=r_{2}e^{\sigma _{2}/2}$ remain invariant. Accordingly, this creation
of the unstable particle should be reflected in the conductance as 
\begin{equation}
G(r_{1},\sigma _{1})=G(r_{2},\sigma _{2})\,\,\,\quad \text{for\quad }%
\,r_{1}e^{\sigma _{1}/2}=r_{2}e^{\sigma _{2}/2}.  \label{M12}
\end{equation}
This means we can control the position of the onset in the conductance by $M$
and its extension in the temperature direction by $\Gamma $. For a model
which possesses unstable particles we found indeed such a behaviour \cite
{CF12}. From the data of the previous subsections we find that the
conductance scales as $G(r_{1},y_{1})=G(r_{2},y_{2})$ for $%
r_{2}y_{1}=r_{1}y_{2}$. Then the comparison with (\ref{M12}) suggests that
we can relate the distance between the two defects to the resonance
parameter as $\sigma =2\ln ($const$/y)$. From the maxima in $|T(\theta )|$
we may identify various $\sigma $s and in fact in this case the net result
can be attributed to two resonances \cite{CF12}.

\subsubsection{Multiple plateaux}

\vspace{-0.2cm} Up to now, we have observed that we always obtain
essentially two plateaux in the conductance, no matter how many ($\geq 2$)
and what type of defects we implement. The natural question arising at this
point is whether it is possible to have a set up which leads to a more
involved plateaux structure? It is clear that if we had many defects in a
row separated far enough from each other such that the relaxation time of
the passing particles is so large that they could be treated as single
rather than multiple defects, then any desired type of multiple plateau
structure could be obtained. In this case the conductance is simply the sum
of the expressions one has for each defect independently. Recalling the
origin of the different plateaux, there is another slightly less obvious
option. The density distribution function $\rho ^{r}$ is a peaked function
of the rapidity and if the resonances in $T^{\mathbf{\alpha }}\left( \theta
\right) $ would be separated far enough, such that they are resolved by $%
\rho ^{r}$, we would also get a multiple plateaux pattern. However, tuning
the distance between the defects or the coupling constant will merely
translate the position of the resonances in the rapidity variable or change
their amplitudes, respectively (see section 2).

\begin{center}
\includegraphics[width=9.cm,height=5.7cm]{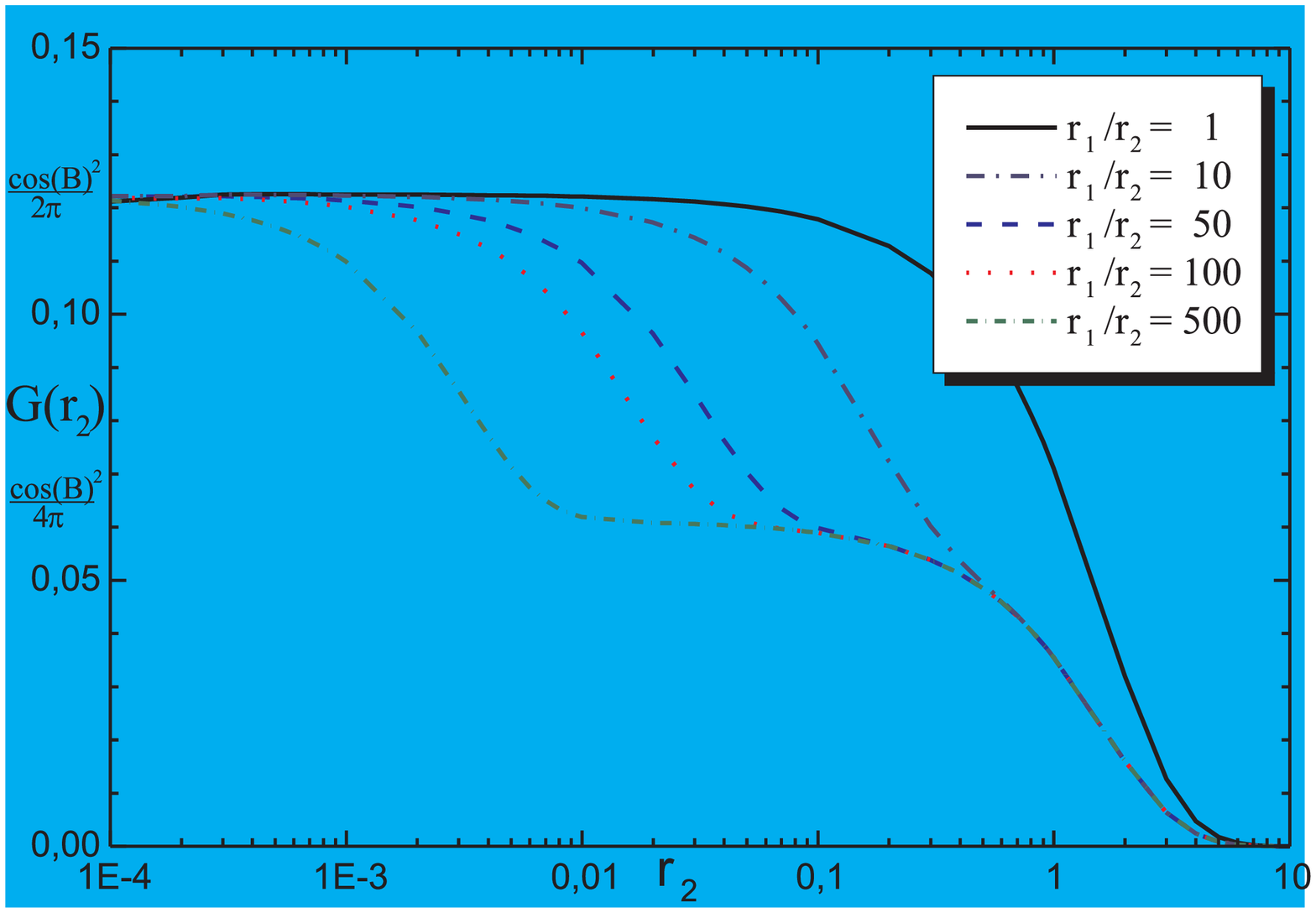}
\end{center}

\noindent {\small Figure 5: Conductance }$G(r_{2})${\small \ for the complex
free Fermion with the energy operator defects as a function of the inverse
temperature }$r_{2},$ {\small for fixed effective coupling constant }$B=0.5$%
{\small \ and varying temperature ratios in the two halves of the wire.}
\medskip

\noindent Therefore the last option left is to change the $\rho ^{r}$s,
which is possible by varying the temperature. Choosing now a configuration
as in figure 3 with different temperatures $T_{1}$ and $T_{2}$, one can
``create'' a second plateau at half the height of the original one. The
reason for this is simply that the cooled half of the wire will cease to
contribute to the conductance as can be directly deduced from (\ref{gg}). We
depict the results of our computations in figure 5.

\noindent From this it also obvious that if we only cool the fraction $x$ of
the wire, the lowest plateau will be positioned at the height $x$ times the
height of the upper plateau. Thus, by combining these different
configurations, i.e., different temperatures or defects, we could produce
any desired plateau structure.

\section{Conductance from the Kubo formula}

\vspace{-0.2cm} \setcounter{equation}{0}

Having computed the DC\ conductance by means of a TBA analysis, we want to
proceed now by introducing an alternative method for the acquisition of the
same quantity, that is the evaluation of the celebrated Kubo formula%
\footnote{%
For a model independent derivation in the context of dynamical respose
theory see, e.g., \cite{KTH}.} \cite{kubo}

\begin{equation}
G(T)=-\lim_{\omega \rightarrow 0}\frac{1}{2\omega \pi ^{2}}%
\int\limits_{-\infty }^{\infty }dt\,\,e^{i\omega t}\,\,\left\langle
J(t)\,J(0)\right\rangle _{T,m}.  \label{kubo}
\end{equation}
The key quantity needed for the explicit computation of (\ref{kubo}) is the
occurrence of the current-current correlation function $\left\langle
J(r)\,J(0)\right\rangle _{T,m}$. In the latter, the subscripts ($T$, $m$)
indicate that, in general, one is interested in a situation when both, the
mass scale of the particles in the quantum field theory and the temperature,
are non-vanishing. This is precisely the same regime in which we have
carried out the TBA analysis in the previous section and ultimately we want
to compare the outcome of both computations. So far, formula (\ref{kubo})
still refers to a situation in which no defect is present in the theory.
Later on we will see how the Kubo formula can also be generalized in order
to incorporate the presence of defects.

As a consequence of the central role played by the two-point function of the
current operator in (\ref{kubo}), we will devote an important part of this
section to recall the key features of a concrete method which will allow for
the computation of such a quantity that is, the form factor bootstrap
approach \cite{KW}. To carry out this program one needs essentially as the
only input the scattering matrix and it is then in principle possible to
compute form factors associated to various local operators of the quantum
field theory under investigation. Form factors are defined as matrix
elements of some local operator $\mathcal{O}(\vec{x})$ located at the origin
between a multiparticle in-state and the vacuum,

\begin{equation}
F_{n}^{\mathcal{O}|\mu _{1}\ldots \mu _{n}}(\theta _{1},\theta _{2}\ldots
,\theta _{n}):=\langle 0|\mathcal{O}(0)|Z_{\mu _{1}}^{\dagger }(\theta
_{1})Z_{\mu _{2}}^{\dagger }(\theta _{2})\ldots Z_{\mu _{n}}^{\dagger
}(\theta _{n})\rangle .  \label{fff}
\end{equation}
They can be obtained by a direct computation once a representation for the
operator involved is known or as solution to a certain set of physically
motivated consistency equations\ \cite{KW, Smir, Zam, BFKZ}, in a similar
fashion as one can determine exact scattering matrices or transmission and
reflection amplitudes for 1+1 dimensional integrable systems as discussed in
section 2.

In the zero-temperature regime, the latter fact is well-known since the
original works \cite{KW} and has lead successfully to the computation of
correlation functions for many models, albeit in most cases only
approximately. It is easy to show that once the corresponding form factors
associated to two local operators $\mathcal{O}$ and $\mathcal{O}^{\prime }$
are known, the computation of their two-point function is reduced to the
task, still non-trivial, of evaluating the following series

\begin{eqnarray}
\left\langle \mathcal{O}(r)\mathcal{O}^{\prime }(0)\right\rangle
_{T=0,m}\!\!\!\! &=&\!\!\!\!\sum\limits_{n=1}^{\infty }\sum\limits_{\mu
_{1}\cdots \mu _{n}}\int \frac{d\theta _{1}\cdots d\theta _{n}}{n!(2\pi )^{n}%
}\prod\limits_{i=1}^{n}e^{-rm_{i}\cosh \theta _{i}}  \nonumber \\
&&\!\!\!\!\quad \quad \times F_{n}^{\mathcal{O}|\mu _{1}\ldots \mu
_{n}}(\theta _{1},\ldots ,\theta _{n})\left[ F_{n}^{\mathcal{O}^{\prime
}|\mu _{1}\ldots \mu _{n}}(\theta _{1},\ldots ,\theta _{n})\right] ^{\ast },
\label{tzero}
\end{eqnarray}
with $x^{\mu }=(-ir,0)$. The previous expression is simply obtained by
introducing a sum over a complete set of states in between the two operators
involved in the correlation function and by shifting the operator $\mathcal{O%
}(r)$ to the origin thereafter. However, as indicated by the subscripts,
this formula applies only to the zero temperature regime. Obviously, when
setting $\mathcal{O=O}^{\prime }=J$, the correlation function (\ref{tzero})
is precisely the quantity entering the Kubo formula (\ref{kubo}), although
for $T=0$. It is therefore necessary to find a generalization of the
expansion (\ref{tzero}) to the $(T\neq 0)$-regime. Such type of
generalization was first suggested in \cite{LLSS} for the Ising model. It
appears, however, to be difficult to generalize this to dynamically
interacting models \cite{CF11} and since by now this has not been achieved
we shall concentrate on the zero temperature regime in this paper.

\subsection{Conductance through an impurity}

\vspace{-0.2cm}

\noindent With the help of (\ref{tzero}) we could in principle compute the
current-current correlation function and therefore evaluate the Kubo formula
when there are no boundaries or defects present. With regard to the
inclusion of boundaries, the first examples in which the Kubo formula was
generalized in order to accommodate that situation were provided in \cite
{LSS}. In there the expression (\ref{kubo}) was evaluated for the
sinh-Gordon and sine-Gordon model in the $m=T=$ $0$\emph{\ }limit and in the
presence of a boundary. This was done by replacing the vacuum state $\left|
0\right\rangle $ with a boundary state $\left| B\right\rangle $ in the
current-current correlation function as follows 
\begin{equation}
\!\left\langle J(r)\,J(0)\right\rangle _{T,m}\!\!\rightarrow \!\left\langle
J(r)\,J(0)B\right\rangle _{T,m}\!\equiv \!\left\langle 0\right| J(r)\,J(0)%
\text{B}\left| 0\right\rangle _{T,m}.  \label{BB}
\end{equation}
The boundary state $\left| B\right\rangle :=$B$\left| 0\right\rangle $ is
understood as the action of a boundary operator B on the vacuum state $%
\left| 0\right\rangle $ \cite{GZ}. Following \cite{GZ}, one exchanges
usually the roles of space and time, such that the correlation functions are
radially rather than time ordered. This is the reason why the boundary
operator B can only enter at the very right or left, since one formulates
such theories in half space. In contrast to the boundary, a defect can also
enter in-between the operators. Therefore, in order to include a defect in (%
\ref{kubo}), one has to consider terms of the form 
\begin{equation}
\left\langle J(r)\,J(0)\right\rangle _{T,m}\rightarrow \left\langle J(r)\,Z_{%
\mathbf{\alpha }}J(0)\right\rangle _{T,m},\quad \left\langle J(r)\,J(0)Z_{%
\mathbf{\alpha }}\right\rangle _{T,m},\quad \left\langle Z_{\mathbf{\alpha }%
}J(r)\,J(0)\right\rangle _{T,m}\,,  \label{DD}
\end{equation}
where $Z_{\mathbf{\alpha }}$ represents the defect operator.

As a consequence of (\ref{DD}), the evaluation of the defect Kubo formula
will require the computation of matrix elements involving the operators $Z_{%
\mathbf{\alpha }}$ 
\begin{equation}
G^{\mathbf{\alpha }}(T)=-\lim_{\omega \rightarrow 0}\frac{1}{2\omega \pi ^{2}%
}\int\limits_{-\infty }^{\infty }dr\,\,e^{i\omega r}\,\,\left\langle
J(r)\,Z_{\alpha _{1}}\cdots Z_{\alpha _{n}}J(0)\right\rangle _{T,m}.
\label{kubod}
\end{equation}
Equation (\ref{kubod}) expresses the conductance for a situation in which $n$
generic defects $Z_{\alpha _{1}}\cdots Z_{\alpha _{n}}$ are present in the
theory and located at positions $y_{\alpha _{1}}\cdots y_{\alpha _{n}}$ in
space. The defect degrees of freedom are encoded into the vector $\mathbf{%
\alpha =}\{\alpha _{1},\cdots ,\alpha _{n}\}$, as done in previous sections.
In order to compare with the TBA results we would like, of course, to
compute the conductance in the massive, finite temperature regime. As
mentioned the evaluation of temperature dependent correlation functions is
still poorly understood, even for the simplest models. In addition, the
presence of the limit in the parameter $\omega $, together with the
introduction of the defect operator $Z_{\mathbf{\alpha }\text{ }}$in the
current-current two-point function makes the generic evaluation of (\ref
{kubod}) fairly involved and constitutes a problem which in general can not
be solved analytically. This is specially cumbersome when double defects are
considered, since the expressions for the reflection and transmission
amplitudes (\ref{tr}), (\ref{tr2}) are, in general, quite messy to handle.
For these reasons it is interesting to start with a more simplified
situation, in which some analytical calculations can still be performed,
that is the $T=0$ regime. One may now view (\ref{DD}) as a three-point
function and extend the expansion (\ref{tzero}) to the case when three
operators enter the correlation function. This will only require the
inclusion of one more set of complete states, such that (\ref{DD}) is
expanded in terms of the form factors of the three operators involved 
\begin{eqnarray}
&&\left\langle J(r)Z_{\mathbf{\alpha }}J(0)\right\rangle
_{T=0,m}\!\!=\!\!\sum\limits_{n,m=1}^{\infty }\sum\limits\Sb \mu _{1}\cdots
\mu _{n}  \\ \nu _{1}\cdots \nu _{m}  \endSb \int \frac{d\theta _{1}\cdots
d\theta _{n}d\tilde{\theta}_{1}\cdots d\tilde{\theta}_{m}}{m!n!(2\pi )^{n+m}}%
F_{n}^{J|\mu _{1}\ldots \mu _{n}}(\theta _{1},\ldots ,\theta _{n})
\label{tree} \\
&&\!\!\times \left\langle Z_{\mu _{n}}(\theta _{n})\ldots Z_{\mu
_{1}}(\theta _{1})|Z_{\mathbf{\alpha }}|Z_{\nu _{1}}(\tilde{\theta}%
_{1})\ldots Z_{\nu _{m}}(\tilde{\theta}_{m})\right\rangle F_{m}^{J|\nu
_{1}\ldots \nu _{m}}(\tilde{\theta}_{1},\ldots ,\tilde{\theta}_{m})^{\ast
}e^{-r\sum\limits_{i=1}^{n}m_{i}\cosh \theta _{i}}.  \nonumber
\end{eqnarray}
We will now restrict ourselves further and consider the massless version of (%
\ref{tree}). In this limit, the results obtained for the conductance should
agree with the UV-limit of the conductance computed by means of (\ref{I11}),
(\ref{gg}). Such a limit can be carried out by exploiting the massless
prescription suggested originally in \cite{massless} and already introduced
in the paragraph before equation (\ref{LR0}). For the form factors in (\ref
{tree}) the massless limit yields 
\begin{equation}
\lim_{\sigma \rightarrow \infty }F_{n}^{\mathcal{O}|\mu _{1}\ldots \mu
_{n}}(\theta _{1}+\eta _{1}\sigma ,\ldots ,\theta _{n}+\eta _{n}\sigma
)=F_{\nu _{1}\cdots \nu _{n}}^{\mathcal{O}|\mu _{1}\ldots \mu _{n}}(\theta
_{1},\ldots ,\theta _{n}),  \label{massff}
\end{equation}
with $\eta _{i}=\pm 1$ and $\nu _{i}=R$ for $\eta _{i}=+$ and $\nu _{i}=L$
for $\eta _{i}=-$. Namely, in the massless limit every massive $n$-particle
form factor is mapped into $2^{n}$ massless form factors. Using these
expressions, performing a Wick rotation and introducing the variable $%
E=\sum_{i=1}^{n}\hat{m}_{i}e^{\theta _{i}}$, we obtain from (\ref{tree}) 
\begin{eqnarray}
&&\left\langle J(r)Z_{\mathbf{\alpha }}J(0)\right\rangle
_{T=m=0}\!\!=\!\!\sum\limits_{n,m=1}^{\infty }\sum\limits\Sb \mu _{1}\cdots
\mu _{n}  \\ \nu _{1}\cdots \nu _{m}  \endSb \int \frac{d\theta _{1}\cdots
d\theta _{n}d\tilde{\theta}_{1}\cdots d\tilde{\theta}_{m}}{m!n!(2\pi )^{n+m}}%
F_{R_{1}\ldots R_{n}}^{J|\mu _{1}\ldots \mu _{n}}(\theta _{1},\ldots ,\theta
_{n})  \label{tree2} \\
&&\!\!\times \left\langle Z_{\mu _{n}}^{R}(\theta _{n})\ldots Z_{\mu
_{1}}^{R}(\theta _{1})|Z_{\mathbf{\alpha }}|Z_{\nu _{1}}^{R}(\tilde{\theta}%
_{1})\ldots Z_{\nu _{m}}^{R}(\tilde{\theta}_{m})\right\rangle F_{R_{1}\ldots
R_{m}}^{J|\nu _{1}\ldots \nu _{m}}(\tilde{\theta}_{1},\ldots ,\tilde{\theta}%
_{m})^{\ast }e^{-irE}.\quad  \nonumber
\end{eqnarray}
We note that for the massless prescription to work, the matrix element
involving the defect $Z_{\mathbf{\alpha }}$ can only depend on the rapidity
differences, which will indeed be the case as we see below. Performing the
variable transformation $\theta _{n}\rightarrow \ln E^{\prime }/\hat{m}%
_{n}-\sum_{i=1}^{n}\hat{m}_{i}/\hat{m}_{n}e^{\theta _{i}}$, we re-write the
r.h.s. of (\ref{tree2}) as 
\begin{eqnarray}
&&\!\!\sum\limits_{n,m=1}^{\infty }\sum\limits\Sb \mu _{1}\cdots \mu _{n} 
\\ \nu _{1}\cdots \nu _{m}  \endSb \int_{0}^{E}dE^{\prime
}\int\limits_{-\infty }^{\ln E^{\prime }/\hat{m}_{n}}\frac{d\theta
_{1}\cdots d\theta _{n-1}}{n!(2\pi )^{n}}\int\limits_{-\infty }^{\infty }%
\frac{d\tilde{\theta}_{1}\cdots d\tilde{\theta}_{m}}{m!(2\pi )^{m}}%
F_{R_{1}\ldots R_{n}}^{J|\mu _{1}\ldots \mu _{n}}(\theta _{1},\ldots ,\theta
_{n}(E^{\prime }))  \label{Hendrix} \\
&&\!\!\times \left\langle Z_{\mu _{n}}^{R}(\theta _{n}(E^{\prime }))\ldots
Z_{\mu _{1}}^{R}(\theta _{1})|Z_{\mathbf{\alpha }}|Z_{\nu _{1}}^{R}(\tilde{%
\theta}_{1})\ldots Z_{\nu _{m}}^{R}(\tilde{\theta}_{m})\right\rangle
F_{R_{1}\ldots R_{m}}^{J|\nu _{1}\ldots \nu _{m}}(\tilde{\theta}_{1},\ldots ,%
\tilde{\theta}_{m})^{\ast }e^{-irE^{\prime }}.\quad  \nonumber
\end{eqnarray}
We substitute now this correlation function into the Kubo formula, shift all
rapidities as $\theta _{i}\rightarrow \theta _{i}+$ $\ln E^{\prime }/\hat{m}%
_{n}$, $\tilde{\theta}_{i}\rightarrow \tilde{\theta}_{i}+$ $\ln E^{\prime }/%
\hat{m}_{n}$ use the Lorentz invariance of the form factors\footnote{%
Denoting by $s$ the Lorentz spin of the operator $\mathcal{O}$ and $\lambda $
being a constant, the form factors satisfy 
\[
F_{n}^{\mathcal{O}|\mu _{1}\ldots \mu _{n}}(\theta _{1}+\lambda ,\ldots
,\theta _{n}+\lambda )=e^{s\lambda }\,F_{n}^{\mathcal{O}|\mu _{1}\ldots \mu
_{n}}(\theta _{1},\ldots ,\theta _{n})\,. 
\]
} and carry out the integration in $dE^{\prime }$%
\begin{eqnarray}
G^{\mathbf{\alpha }}\!\!\! &=&\!\!\!-\lim_{\omega \rightarrow 0}\frac{\omega
^{2s-2}}{\hat{m}_{n}^{2s}\pi }\sum\limits\Sb \mu _{1}\cdots \mu _{n}  \\ \nu
_{1}\cdots \nu _{m}  \endSb \int\limits_{-\infty }^{0}\frac{d\theta
_{1}\cdots d\theta _{n-1}}{n!(2\pi )^{n}}\int\limits_{-\infty }^{\infty }%
\frac{d\tilde{\theta}_{1}\cdots d\tilde{\theta}_{m}}{m!(2\pi )^{m}}\frac{1}{%
1-\sum_{i=1}^{n-1}\hat{m}_{i}/\hat{m}_{n}e^{\theta _{i}}}  \nonumber \\
&&\times \left\langle Z_{\mu _{n}}^{R}(\ln (1-\sum\nolimits_{i=1}^{n-1}\hat{m%
}_{i}/\hat{m}_{n}e^{\theta _{i}}))\ldots Z_{\mu _{1}}^{R}(\theta _{1})|Z_{%
\mathbf{\alpha }}|Z_{\nu _{1}}^{R}(\tilde{\theta}_{1})\ldots Z_{\nu
_{m}}^{R}(\tilde{\theta}_{m})\right\rangle \quad  \label{conn} \\
&&\times F_{R_{1}\ldots R_{n}}^{J|\mu _{1}\ldots \mu _{n}}(\theta
_{1},\ldots ,\ln (1-\sum_{i=1}^{n-1}\hat{m}_{i}/\hat{m}_{n}e^{\theta
_{i}}))F_{R_{1}\ldots R_{m}}^{J|\nu _{1}\ldots \nu _{m}}(\tilde{\theta}%
_{1},\ldots ,\tilde{\theta}_{m})^{\ast }\,\,.  \nonumber
\end{eqnarray}
We state various observations: Since the matrix element involving the defect
only depends on the rapidity difference, it is not affected by the shifts.
The Lorentz spin $s=1$ plays a very special role in (\ref{conn}), which
makes the current operator especially distinguished. In that case the r.h.s.
of (\ref{conn}) becomes independent of the frequency $\omega $ and the limit
is carried out trivially. Furthermore, since the final expression has to be
independent of $\hat{m}_{n}$, we deduce that the form factors have to be
linearly dependent on $\hat{m}_{n}$.

One may now compute the form factors by solving either the associated
consistency equations or by using concrete realizations of the operators.
For those form factors involving the current operator $J$, a realization in
terms of the ZF-algebra was given in \cite{Fform} for complex free Fermion
type models and used to compute the corresponding matrix elements. We will
determine form factors involving the defect operator in the same fashion,
which means we require a concrete realization for the operator $Z_{\alpha }$.

\subsection{Realization of the defect operator}

\vspace{-0.2cm}

\noindent A realization of $Z_{\alpha }$ can be achieved very much in
analogy to a realization of local operators, i.e. as exponentials of
bilinears in the ZF-operators \cite{SMJ}. For the case of a boundary a
generic model independent realization for the boundary operator B was
originally proposed in \cite{GZ} for the parity invariant case, i.e., $R=%
\tilde{R}$ . This proposal was generalized to the defect operator in \cite
{Konik2} with the same restriction and for self-conjugated particles. This
realization was used by the authors for the computation of various matrix
elements involving the defect operator. Our aim in this section is to extend
this realization in order to incorporate the possibility of parity breaking
as well as non self-conjugated particles. A non-trivial consistency check
for the validity of our proposal will be ultimately provided when exploiting
it in the computation of the conductance, obtained before by entirely
different means, that is the TBA approach. The realization we want to
propose here is a direct generalization of the one presented in \cite{Konik2}%
, namely 
\begin{equation}
Z_{\mathbf{\alpha }}=:\exp [\frac{1}{4\pi }\int\nolimits_{-\infty }^{\infty
}D_{\mathbf{\alpha }}(\theta )\,d\theta ]:\text{ },  \label{D}
\end{equation}
where\ : \ \ : denotes normal ordering and the operator $D_{\alpha }(\theta
) $ has the form 
\begin{eqnarray}
D_{\mathbf{\alpha }}(\theta )\!\!\! &=&\!\!\!\sum\limits_{i}\left[ K_{i}^{%
\mathbf{\alpha }}(\theta )Z_{i}^{\dagger }(\theta )Z_{\bar{\imath}}^{\dagger
}(-\theta )+\tilde{K}_{i}^{\mathbf{\alpha }}(\theta )^{\ast }\,Z_{\bar{\imath%
}}(-\theta )Z_{i}(\theta )\right.  \nonumber \\
&&\left. \qquad +W_{i}^{\mathbf{\alpha }}(\theta )Z_{i}^{\dagger }(\theta
)Z_{i}(\theta )+\tilde{W}_{i}^{\mathbf{\alpha }}(\theta )^{\ast
}Z_{i}^{\dagger }(-\theta )Z_{i}(-\theta )\right] ,  \label{gend}
\end{eqnarray}
with 
\begin{eqnarray}
&&K_{i}^{\mathbf{\alpha }}(\theta ):=R_{i}^{\mathbf{\alpha }}(\frac{i\pi }{2}%
-\theta ),\quad \qquad \tilde{K}_{i}^{\mathbf{\alpha }}(\theta ):=\tilde{R}%
_{i}^{\mathbf{\alpha }}(\frac{i\pi }{2}-\theta ),  \label{s1} \\
&&W_{i}^{\mathbf{\alpha }}(\theta ):=T_{i}^{\mathbf{\alpha }}(\frac{i\pi }{2}%
-\theta ),\quad \qquad \tilde{W}_{i}^{\mathbf{\alpha }}(\theta ):=\tilde{T}%
_{i}^{\mathbf{\alpha }}(\frac{i\pi }{2}-\theta ).  \label{s2}
\end{eqnarray}
In comparison with \cite{Konik2} we have used a slightly different
normalization factor, since in general we have contributions in the sum over 
$i$ in (\ref{gend}) including both particles and anti-particles, as for the
complex free Fermion we shall treat below. Following the arguments given in 
\cite{GZ}, the operator $D_{\mathbf{\alpha }}(\theta )$ depends on the
amplitudes $R(\theta )$, $T(\theta )$, $\tilde{R}(\theta )$ and $\tilde{T}%
(\theta )$ with their arguments shifted according to (\ref{s1})-(\ref{s2}),
as considered also in \cite{DMS,Konik2}. The reason for these shifts is the
exchange of the roles played by the space and time coordinates $x^{\mu
}=(t,x)\rightarrow (ix,it)$, which was already mentioned after (\ref{BB}).
Doing this and keeping simultaneously the product $\ x^{\mu }\cdot p_{\mu }$
invariant requires the rapidity shifts in (\ref{s1})-(\ref{s2}). In our
context, this implies that we must now not only perform the shifts (\ref{s1}%
)-(\ref{s2}) in our expressions, but also, with regard to the positions of
the defects, the change $y_{\alpha }\rightarrow iy_{\alpha }$ should be
implemented. The latter replacement will play an important role since,
similar as in the TBA case, the amplitudes (\ref{s1})-(\ref{s2}) will become
in this way strongly oscillating functions of $y_{\alpha }$. Therefore, we
may be able to carry out once more certain analytical calculations, by
replacing the mentioned functions with their mean values. In (\ref{gend}) we
have already specialized to the case when the reflection and transmission
amplitudes are diagonal both with respect to the particle and defect degrees
of freedom, since that will be the situation for all the examples we want to
treat in this paper.

\subsection{Defect matrix elements}

\vspace{-0.2cm} \noindent Having now a concrete generic realization of the
defect (\ref{gend}), we can compute the defect matrix elements. One way of
doing this is to solve a set of consistency equations which relate the lower
particle matrix elements to higher particle ones, similar as in the standard
form factor program \cite{KW}. Such kind of iterative equations were
proposed in \cite{DMS} for a parity invariant defect and for a real free
fermionic and bosonic theory. First we note that the operator (\ref{D})
becomes 
\begin{equation}
Z_{\mathbf{\alpha }}=:\exp [\frac{1}{2\pi }\int\nolimits_{-\infty }^{\infty
}d\theta \sum\limits_{i}Z_{i}^{\dagger }(\theta )Z_{i}(\theta )\,]:,
\end{equation}
in the limit $R=\tilde{R}=0$ and $T=\tilde{T}=1$. The defect should act in
this case as the identity operator and, according to (\ref{Z1/2}), 
\begin{equation}
\langle Z_{i}(\theta _{1})Z_{\mathbf{\alpha }}Z_{j}^{\dagger }(\theta
_{2})\rangle =2\pi \,\delta (\theta _{12})\delta _{ij},
\end{equation}
holds, simply by employing Wick's theorem when carrying out the necessary
contractions. For two particles we find, 
\begin{eqnarray}
\left\langle Z_{\bar{\imath}}(\theta _{1})Z_{i}(\theta _{2})Z_{\mathbf{%
\alpha }}\right\rangle &=&\pi \hat{K}_{i}^{\mathbf{\alpha }}(\theta
_{2})\delta (\hat{\theta}_{12}),  \label{i1} \\
\langle Z_{\mathbf{\alpha }}Z_{i}^{\dagger }(\theta _{1})Z_{\bar{\imath}%
}^{\dagger }(\theta _{2})\rangle &=&\pi \,\hat{K}_{i}^{\mathbf{\alpha }%
}(\theta _{1})^{\ast }\delta (\hat{\theta}_{12}),  \label{i2} \\
\langle Z_{i}(\theta _{1})Z_{\mathbf{\alpha }}Z_{j}^{\dagger }(\theta
_{2})\rangle &=&\pi \,\hat{W}_{i}^{\mathbf{\alpha }}(\theta _{1})\delta
(\theta _{12})\delta _{ij},  \label{i3}
\end{eqnarray}
where we recall from section 2 the notation$\ \hat{\theta}_{12}=\theta
_{1}+\theta _{2}$ and $\theta _{12}=\theta _{1}-\theta _{2}$. For later
convenience we have introduced the functions 
\begin{eqnarray}
\hat{K}_{i}^{\mathbf{\alpha }}(\theta ) &=&K_{i}^{\mathbf{\alpha }}(\theta
)\,+S_{\bar{\imath}i}(-2\theta )K_{\bar{\imath}}^{\mathbf{\alpha }}(-\theta
)=\tilde{K}_{i}^{\mathbf{\alpha }}(\theta )\,+S_{i\bar{\imath}}(2\theta )%
\tilde{K}_{\bar{\imath}}^{\mathbf{\alpha }}(-\theta ),  \label{uss1} \\
\hat{W}_{i}^{\mathbf{\alpha }}(\theta ) &=&W_{i}^{\mathbf{\alpha }}(\theta
)\,+\tilde{W}_{i}^{\mathbf{\alpha }}(-\theta )^{\ast }=\tilde{W}_{\bar{\imath%
}}^{\mathbf{\alpha }}(-\theta )\,+W_{\bar{\imath}}^{\mathbf{\alpha }}(\theta
)^{\ast }=\hat{W}_{\bar{\imath}}^{\mathbf{\alpha }}(\theta )^{\ast },
\label{uss2}
\end{eqnarray}
since the $K_{i}^{\mathbf{\alpha }}$, $\tilde{K}_{i}^{\mathbf{\alpha }}$,$%
W_{i}^{\mathbf{\alpha }}\,$and $\tilde{W}_{i}^{\mathbf{\alpha }}$ amplitudes
defined by (\ref{s1})-(\ref{s2}) will repeatedly appear in the combinations (%
\ref{uss1}), (\ref{uss2}) in what follows. The latter equalities in (\ref
{uss1}), (\ref{uss2}) follow simply from 
\begin{equation}
\tilde{W}_{i}^{\mathbf{\alpha }}(\theta )=W_{\bar{\imath}}^{\mathbf{\alpha }%
}(-\theta )=\tilde{W}_{\bar{\imath}}^{\mathbf{\alpha }}(i\pi -\theta )^{\ast
},\text{\thinspace \thinspace }\tilde{K}_{i}^{\mathbf{\alpha }}(\theta )=S_{i%
\bar{\imath}}(2\theta )K_{\bar{\imath}}^{\mathbf{\alpha }}(-\theta )=S_{i%
\bar{\imath}}(2\theta )\tilde{K}_{\bar{\imath}}^{\mathbf{\alpha }}(i\pi
-\theta )^{\ast },  \label{ck}
\end{equation}
which are in turn consequences of the crossing-hermiticity properties (\ref
{c1})-(\ref{c2}). Having these matrix elements we can construct the ones
involving more particles recursively from 
\begin{eqnarray}
&&F_{\mathbf{\alpha }}^{\mu _{m}\ldots \mu _{1}\nu _{1}\ldots \nu
_{n}}(\theta _{m}\ldots \theta _{1},\theta _{1}^{\prime }\ldots \theta
_{n}^{\prime }):=\left\langle Z_{\mu _{m}}(\theta _{m})\,\ldots Z_{\mu
_{1}}(\theta _{1})Z_{\mathbf{\alpha }}\,Z_{\nu _{1}}^{\dagger }(\theta
_{1}^{\prime })\ldots Z_{\nu _{n}}^{\dagger }(\theta _{n}^{\prime
})\right\rangle =\quad  \nonumber \\
&&\quad \pi \sum_{l=2}^{m}\delta _{\mu _{1}\bar{\mu}_{l}}\delta (\hat{\theta}%
_{1l})\hat{K}_{\mu _{1}}^{\mathbf{\alpha }}(\theta
_{1})\prod_{p=1}^{l-1}S_{\mu _{1}\mu _{p}}(\theta _{1p})F_{\mathbf{\alpha }%
}^{\mu _{m}\ldots \check{\mu}_{l}\ldots \mu _{2}\nu _{1}\ldots \nu
_{n}}(\theta _{m}\ldots \check{\theta}_{l}\ldots \theta _{2},\theta
_{1}^{\prime }\ldots \theta _{n}^{\prime })  \label{MM} \\
&&+\pi \sum_{l=1}^{n}\delta _{\mu _{1}\nu _{l}}\delta (\theta _{1}-\theta
_{l}^{\prime })\hat{W}_{\mu _{1}}^{\mathbf{\alpha }}(\theta
_{1})\prod_{p=1}^{l-1}S_{\mu _{1}\nu _{p}}(\theta _{1p})F_{\mathbf{\alpha }%
}^{\mu _{m}\ldots \mu _{2}\nu _{1}\ldots \check{\nu}_{l}\ldots \nu
_{n}}(\theta _{m}\ldots \theta _{2},\theta _{1}^{\prime }\ldots \check{\theta%
}_{l}^{\prime }\ldots \theta _{n}^{\prime })  \nonumber
\end{eqnarray}
\begin{eqnarray}
&&F_{\mathbf{\alpha }}^{\mu _{m}\ldots \mu _{1}\nu _{1}\ldots \nu
_{n}}(\theta _{m}\ldots \theta _{1},\theta _{1}^{\prime }\ldots \theta
_{n}^{\prime })=  \label{MM2} \\
&&\quad \pi \sum_{l=2}^{n}\delta _{\nu _{1}\bar{\nu}_{l}}\delta (\hat{\theta}%
_{1l}^{\prime })\hat{K}_{\nu _{1}}^{\mathbf{\alpha }}(\theta _{1}^{\prime
})^{\ast }\prod_{p=1}^{l-1}S_{\nu _{1}\mu _{p}}(\theta _{1p})F_{\mathbf{%
\alpha }}^{\mu _{m}\ldots \mu _{1}\nu _{2}\ldots \check{\nu}_{l}\ldots \nu
_{n}}(\theta _{m}\ldots \theta _{1},\theta _{2}^{\prime }\ldots \check{\theta%
}_{l}^{\prime }\ldots \theta _{n}^{\prime })  \nonumber \\
&&+\pi \sum_{l=1}^{m}\delta _{\nu _{1}\mu _{l}}\delta (\theta _{1}^{\prime
}-\theta _{l})\hat{W}_{\nu _{1}}^{\mathbf{\alpha }}(\theta _{1}^{\prime
})^{\ast }\prod_{p=1}^{l-1}S_{\nu _{1}\mu _{p}}(\theta _{1p})F_{\mathbf{%
\alpha }}^{\mu _{m}\ldots \check{\mu}_{l}\ldots \mu _{1}\nu _{2}\ldots \nu
_{n}}(\theta _{m}\ldots \check{\theta}_{l}\ldots \theta _{1},\theta
_{2}^{\prime }\ldots \theta _{n}^{\prime })  \nonumber
\end{eqnarray}
Here we denoted with the check on the rapidities $\check{\theta}$ the
absence of the corresponding particle in the matrix element. It is clear
from the expressions (\ref{D}) and (\ref{gend}) that the only possible
non-vanishing matrix elements (\ref{MM}) are those when $n+m$ is even.
Taking (\ref{i1})-(\ref{i3}) as the initial conditions for the recursive
equation (\ref{MM})-(\ref{MM2}), we can now either solve them iteratively or
use (\ref{D}) and evaluate the matrix elements directly.

\subsection{Free Fermion with defects}

\vspace{-0.2cm}Similar as for the TBA we want to exemplify our general
formulae with the complex free Fermion. We consider now the
particularization of the defect realization (\ref{gend}) to this case. Then
the generators of the ZF-algebra \emph{Z}$_{i}(\theta ),$ \emph{Z}$%
_{i}^{\dagger }(\theta )$ are just the usual creation and annihilation
operators $a_{i}(\theta ),\,a_{i}^{\dagger }(\theta )$ in the free fermionic
Fock space and we have to distinguish between particles \emph{i }and
antiparticles \emph{\={\i}}. For the complex free Fermion it is interesting
to notice that the realization (\ref{D}) resembles very much the one
employed in \cite{SMJ,Fform} for a prototype local field.

\subsubsection{Defect matrix elements}

\vspace{-0.2cm}Let us now use (\ref{D})-(\ref{gend}) in order to evaluate
matrix elements involving the defect operator. In what follows, the most
relevant matrix elements are those involving four particles, for which we
compute 
\begin{eqnarray}
&&\langle a_{i}(\theta _{1})\,a_{\bar{\imath}}(\theta _{2})Z_{\mathbf{\alpha 
}}\,a_{\bar{\imath}}^{\dagger }(\theta _{3})\,a_{i}^{\dagger }(\theta
_{4})\rangle =w_{i\bar{\imath}}^{\mathbf{\alpha }}(\theta _{1,}\theta
_{2})\delta (\theta _{14})\delta (\theta _{23})+k_{ii}^{\mathbf{\alpha }%
}(\theta _{1,}\theta _{4})\delta (\hat{\theta}_{12})\delta (\hat{\theta}%
_{34}),\qquad  \label{4p} \\
&&\langle a_{i}(\theta _{1})\,a_{i}(\theta _{2})Z_{\mathbf{\alpha }%
}\,a_{j}^{\dagger }(\theta _{3})\,a_{j}^{\dagger }(\theta _{4})\rangle =-\pi
^{2}\hat{W}_{i}^{\mathbf{\alpha }}(\theta _{1})\hat{W}_{i}^{\mathbf{\alpha }%
}(\theta _{2})\delta (\theta _{13})\delta (\theta _{24})\delta _{ij}, \\
&&\langle a_{i}(\theta _{1})a_{k}(\theta _{2})a_{i}(\theta _{3})Z_{\mathbf{%
\alpha }}a_{i}^{\dagger }(\theta _{4})\rangle =\pi ^{2}\hat{W}_{i}^{\mathbf{%
\alpha }}(\theta _{4})\hat{K}_{i}^{\mathbf{\alpha }}(-\theta _{2})\left[
\delta (\theta _{14})\delta (\hat{\theta}_{23})-\delta (\hat{\theta}%
_{12})\delta (\theta _{34})\right] \delta _{i\bar{k}},  \nonumber \\
&&\langle a_{i}(\theta _{1})Z_{\mathbf{\alpha }}a_{i}^{\dagger }(\theta
_{2})a_{k}^{\dagger }(\theta _{3})a_{i}^{\dagger }(\theta _{4})\rangle =\pi
^{2}\hat{W}_{i}^{\mathbf{\alpha }}(\theta _{1})\hat{K}_{i}^{\mathbf{\alpha }%
}(-\theta _{3})^{\ast }\left[ \delta (\hat{\theta}_{23})\delta (\theta
_{14})-\delta (\theta _{12})\delta (\hat{\theta}_{34})\right] \delta _{i\bar{%
k}},  \nonumber
\end{eqnarray}
with the abbreviations 
\begin{equation}
w_{i\bar{\imath}}^{\mathbf{\alpha }}(\theta _{1,}\theta _{2})=\pi ^{2}\hat{W}%
_{i}^{\mathbf{\alpha }}(\theta _{1})\hat{W}_{\bar{\imath}}^{\mathbf{\alpha }%
}(\theta _{2})\qquad k_{ii}^{\mathbf{\alpha }}(\theta _{1,}\theta _{2})=\pi
^{2}\hat{K}_{i}^{\mathbf{\alpha }}(\theta _{1})\hat{K}_{i}^{\mathbf{\alpha }%
}(\theta _{2})^{\ast }\,.  \label{222}
\end{equation}
One can also find solutions for all $n$-particle form factors either from (%
\ref{MM})-(\ref{MM2}) or by direct computation. For instance we compute 
\begin{eqnarray}
&&F_{\mathbf{\alpha }}^{m\times (i\bar{\imath})\,n\times (\bar{\imath}%
i)}(\theta _{2m}\ldots \theta _{1},\theta _{1}^{\prime }\ldots \theta
_{2n}^{\prime })=\sum_{k=0}^{\min (n,m)}\frac{(-1)^{m+n-2k}\pi ^{n+m}}{%
(m-k)!(n-k)!k!k!}\int\nolimits_{-\infty }^{\infty }d\beta _{1}\ldots d\beta
_{2n+2m}\,\,  \nonumber \\
&&\times \det \mathcal{A}^{2n}(\beta _{1}\ldots \beta _{2n};\theta
_{1}^{\prime }\ldots \theta _{2n}^{\prime })\det \mathcal{A}^{2m}(\beta
_{2n+1}\ldots \beta _{2n+2m};\theta _{1}\ldots \theta _{2m})  \nonumber \\
&&\times \prod_{p=1}^{k}\hat{W}_{i}^{\mathbf{\alpha }}(\beta _{2p})\hat{W}_{%
\bar{\imath}}^{\mathbf{\alpha }}(\beta _{2p-1})\delta (\beta _{2p}-\beta
_{2n+2p})\delta (\beta _{2p-1}-\beta _{2n+2p-1})  \label{even} \\
&&\times \prod_{p=1+k}^{n}\hat{K}_{i}^{\mathbf{\alpha }}(\beta _{2p})^{\ast
}\delta (\beta _{2p}+\beta _{2p-1})\prod_{p=1+k+n}^{n+m}\hat{K}_{i}^{\mathbf{%
\alpha }}(\beta _{2p})\delta (\beta _{2p}+\beta _{2p-1})\,,  \nonumber
\end{eqnarray}
where $\mathcal{A}^{\ell }(\theta _{1}\ldots \theta _{\ell };\theta
_{1}^{\prime }\ldots \theta _{\ell }^{\prime })$ is a rank $\ell $ matrix
whose entries are given by 
\begin{equation}
\mathcal{A}_{ij}^{\ell }=\cos ^{2}[(i-j)\pi /2]\delta (\theta _{i}-\theta
_{j}^{\prime })\,,\qquad \,\quad 1\leq i,j\leq \ell \,.
\end{equation}
The matrix elements are computed similarly as in \cite{Fform} and references
therein. Likewise we compute 
\begin{eqnarray}
F_{\mathbf{\alpha }}^{n\times i+m\times i}(\theta _{n}\ldots \theta
_{1},\theta _{1}^{\prime }\ldots \theta _{m}^{\prime }) &=&\delta _{n,m}%
\frac{\pi ^{n}(-1)^{n-1}}{n!}\int\nolimits_{-\infty }^{\infty }d\beta
_{1}\ldots d\beta _{n}\prod\limits_{k=1}^{n}\hat{W}_{i}^{\mathbf{\alpha }%
}(\theta _{k})  \nonumber \\
&&\!\!\!\!\!\!\!\!\!\!\!\!\!\!\!\!\!\!\!\!\!\!\!\!\!\!\!\!\!\!\!\!\!\!\!\!\!%
\!\!\!\!\!\!\!\!\!\!\!\!\!\!\!\!\!\!\times \det \mathcal{B}^{n}(\theta
_{n}\ldots \theta _{1};\beta _{1}\ldots \beta _{n})\det \mathcal{B}%
^{n}(\beta _{1}\ldots \beta _{n};\theta _{1}^{\prime }\ldots \theta
_{n}^{\prime }),  \label{allis}
\end{eqnarray}
where we introduced a new rank $\ell $ matrix $\mathcal{B}^{\ell }(\theta
_{1}\ldots \theta _{\ell };\theta _{1}^{\prime }\ldots \theta _{\ell
}^{\prime })$ whose entries are now simply given by 
\begin{equation}
\mathcal{B}_{ij}^{\ell }=\delta (\theta _{i}-\theta _{j}^{\prime }),\qquad
\,\quad 1\leq i,j\leq \ell \,.
\end{equation}
Since (\ref{allis}) is simpler than (\ref{even}) we use it to demonstrate
explicitly that it satisfies the recurrence relations (\ref{MM}) and (\ref
{MM2}). The other cases work the same way. Starting with the expansion of
the determinant $\det \mathcal{B}^{n}(\theta _{n}\ldots \theta _{1};\beta
_{1}\ldots \beta _{n})$ with respect to the row involving the variable $%
\theta _{1}$ gives 
\begin{equation}
\det \mathcal{B}^{n}(\theta _{n}\ldots \theta _{1};\beta _{1}\ldots \beta
_{n})\!=\!\sum\limits_{l=1}^{n}(-1)^{n+l+1}\delta (\theta _{1}-\beta
_{l})\det \mathcal{B}^{n-1}(\theta _{n}\ldots \theta _{2};\beta _{1}\ldots 
\check{\beta}_{l}\ldots \beta _{n}).\!  \label{bob}
\end{equation}
Inserting then (\ref{bob}) into (\ref{allis}), we obtain 
\begin{eqnarray}
&&F_{\mathbf{\alpha }}^{n\times i+m\times i}(\theta _{n}\ldots \theta
_{1},\theta _{1}^{\prime }\ldots \theta _{m}^{\prime })=\delta _{n,m}\frac{%
\pi ^{n}}{n!}\sum\limits_{l=1}^{n}\hat{W}_{i}^{\mathbf{\alpha }}(\theta
_{1})\int\nolimits_{-\infty }^{\infty }d\beta _{1}\ldots d\check{\beta}%
_{l}\ldots d\beta _{n}  \nonumber \\
&&\times (-1)^{l}\prod\limits_{k=1}^{l-1}\hat{W}_{i}^{\mathbf{\alpha }%
}(\theta _{k})\prod\limits_{k=l+1}^{n}\hat{W}_{i}^{\mathbf{\alpha }}(\theta
_{k})\det \mathcal{B}^{n-1}(\theta _{n}\ldots \theta _{2};\beta _{1}\ldots 
\check{\beta}_{l}\ldots \beta _{n})  \nonumber \\
&&\times \det \mathcal{B}^{n}(\beta _{1}\ldots \beta _{l}\rightarrow \theta
_{1}\ldots \beta _{n};\theta _{1}^{\prime }\ldots \theta _{n}^{\prime }).
\label{koelln}
\end{eqnarray}
Expanding now the second determinant in (\ref{koelln}) with respect to the $%
l $-th row, which involves the rapidity $\theta _{1}$, and using the fact
that the $\beta $s are just integration variables and therefore, the sum in $%
l$ gives actually $n$ times the same contribution, we can write 
\begin{eqnarray}
&&F_{\mathbf{\alpha }}^{n\times i+m\times i}(\theta _{n}\ldots \theta
_{1},\theta _{1}^{\prime }\ldots \theta _{m}^{\prime })=\delta _{n,m}\frac{%
\pi ^{n}}{(n-1)!}\sum\limits_{p=1}^{n}\hat{W}_{i}^{\mathbf{\alpha }}(\theta
_{1})\delta (\theta _{1}-\theta _{p}^{\prime })\int\nolimits_{-\infty
}^{\infty }d\beta _{1}\ldots d\beta _{n-1}  \nonumber \\
&&\times (-1)^{p}\prod\limits_{k=1}^{p-1}\hat{W}_{i}^{\mathbf{\alpha }%
}(\theta _{k})\prod\limits_{k=p+1}^{n}\hat{W}_{i}^{\mathbf{\alpha }}(\theta
_{k})\det \mathcal{B}^{n-1}(\theta _{n}\ldots \theta _{2};\beta _{1}\ldots
\beta _{n-1})  \nonumber \\
&&\times \det \mathcal{B}^{n-1}(\beta _{1}\ldots \beta _{n-1};\theta
_{1}^{\prime }\ldots \check{\theta}_{p}^{\prime }\ldots \theta _{n}^{\prime
}).  \label{lt}
\end{eqnarray}
We recognize now the matrix element with two particles less on the l.h.s. of
(\ref{lt}) and can re-write it as 
\begin{eqnarray}
F_{\mathbf{\alpha }}^{n\times i+n\times i}(\theta _{n}\ldots \theta
_{1},\theta _{1}^{\prime }\ldots \theta _{n}^{\prime }) &=&\pi
\sum\limits_{p=1}^{n}(-1)^{p-1}\hat{W}_{i}^{\mathbf{\alpha }}(\theta
_{1})\delta (\theta _{1}-\theta _{p}^{\prime })  \label{brilliant} \\
&&\times F_{\mathbf{\alpha }}^{(n-1)\times i+(n-1)\times i}(\theta
_{n-1}\ldots \theta _{2},\theta _{1}^{\prime }\ldots \check{\theta}%
_{p}^{\prime }\ldots \theta _{n}^{\prime }),  \nonumber
\end{eqnarray}
which is in complete agreement with (\ref{MM}). The validity of (\ref{MM2})
can be checked similarly.

\subsubsection{Conductance in the $T=m=0$ regime}

\vspace{-0.2cm} \noindent It is well-known that for a free Fermion theory
(also for a single complex free Fermion) the conformal $U(1)$%
-current-current correlation function is simply 
\begin{equation}
\left\langle J(r)J(0)\right\rangle _{T=m=0}=\frac{1}{r^{2}}.  \label{mt0}
\end{equation}
This expression can also be obtained by using the expansion (\ref{tzero}),
together with the massless prescription outlined before (\ref{LR0}) (see 
\cite{CF11}) and the expressions for the only non-vanishing form factors of
the current operator in the complex free Fermion theory 
\begin{equation}
F_{2}^{J|\bar{\imath}i}(\theta ,\tilde{\theta})=-F_{2}^{J|i\bar{\imath}%
}(\theta ,\tilde{\theta})=-i\pi me^{\frac{\theta +\tilde{\theta}}{2}}\,.
\label{currentff}
\end{equation}
In particular, the massless limit of the previous expressions gives,
according to the massless prescription, 
\begin{eqnarray}
F_{RR}^{J|\bar{\imath}i}(\theta ,\tilde{\theta}) &=&-F_{RR}^{J|i\bar{\imath}%
}(\theta ,\tilde{\theta})=-2\pi i\,\hat{m}e^{\frac{\theta +\tilde{\theta}}{2}%
}\,,  \label{mf1} \\
F_{LL}^{J|\bar{\imath}i}(\theta ,\tilde{\theta}) &=&F_{LR}^{J|\bar{\imath}%
i}(\theta ,\tilde{\theta})=F_{RL}^{J|\bar{\imath}i}(\theta ,\tilde{\theta}%
)=0,  \label{mf2} \\
F_{LL}^{J|i\bar{\imath}}(\theta ,\tilde{\theta}) &=&F_{LR}^{J|i\bar{\imath}%
}(\theta ,\tilde{\theta})=F_{RL}^{J|i\bar{\imath}}(\theta ,\tilde{\theta}%
)=0\,.  \label{:->}
\end{eqnarray}
Inserting (\ref{mt0}) into (\ref{kubo}) reduces the problem of finding the
Fourier transform of the function $r^{-2}$ which is given by $\mathcal{P}%
\int_{-\infty }^{\infty }dr\,\,e^{i\omega r}r^{-2}=-\pi \omega $ for $\omega
>0$, with $\mathcal{P}$ denoting the principle value. This yields in the
absence of a defect $G^{0}(0)=1/2\pi $, in complete agreement with the limit
(\ref{moorhuhn}).

Let us now consider a more complicated situation, that is, the evaluation of
(\ref{kubod}) for $T=m=0$ in the presence of $n$ defects $\,Z_{\alpha
_{1}}\cdots Z_{\alpha _{n}}$ located at positions $y_{\alpha _{1}}\ldots
y_{\alpha _{n}}$ in space. The correlation function (\ref{DD}) can now be
obtained with the help of (\ref{tzero}), which has to be generalized for
three-point functions. This requires the inclusion of one more sum over a
complete set of states in (\ref{tzero}). Fortunately the only non-vanishing
form factors of the current are (\ref{mf1}), which means the expansion (\ref
{tzero}) will already terminate for two particles. Explicitly, we find 
\begin{eqnarray}
&&\left\langle J(r)Z_{\alpha _{1}}\cdots Z_{\alpha _{n}}J(0)\right\rangle
_{T=m=0}=\sum\limits_{i}\int\limits_{-\infty }^{\infty }\frac{d\theta
_{1}d\theta _{2}d\theta _{3}d\theta _{4}}{2(2\pi )^{4}}F_{RR}^{J|\bar{\imath}%
i}(\theta _{1},\theta _{2})\left[ F_{RR}^{J|\bar{\imath}i}(\theta
_{3},\theta _{4})\right] ^{\ast }\,\,\,\quad \quad  \nonumber \\
&&\qquad \qquad \qquad \qquad \times e^{-r\hat{m}(e^{\theta _{1}}+e^{\theta
_{2}})}\langle a_{i}(\theta _{1})a_{\bar{\imath}}(\theta _{2})Z_{\alpha
_{1}}\cdots Z_{\alpha _{n}}a_{\bar{\imath}}^{\dagger }(\theta
_{3})a_{i}^{\dagger }(\theta _{4})\rangle _{m=0},\qquad \quad  \label{jzj}
\end{eqnarray}
In the light of the expressions (\ref{222}), we can re-write (\ref{jzj}) in
a more explicit form without the need of specifying a concrete defect yet.
Inserting (\ref{4p}) and (\ref{mf1}) into (\ref{jzj}), we find 
\begin{eqnarray}
&&\left\langle J(r)Z_{\alpha _{1}}\cdots Z_{\alpha _{n}}J(0)\right\rangle
_{T=m=0}=\frac{\hat{m}^{2}}{2}\sum\limits_{i}\left[ \int\limits_{-\infty
}^{\infty }\frac{d\theta _{1}}{2}e^{-2r\hat{m}\cosh \theta _{1}}\hat{K}_{i}^{%
\mathbf{\alpha |}R}(\theta _{1})\int\limits_{-\infty }^{\infty }\frac{%
d\theta _{2}}{2}\hat{K}_{i}^{\mathbf{\alpha |}R}(\theta _{2})^{\ast }\right.
\nonumber \\
&&\left. \qquad +\int\limits_{-\infty }^{\infty }\frac{d\theta _{1}}{2}%
e^{\theta _{1}-r\hat{m}e^{\theta _{1}}}\hat{W}_{i}^{\mathbf{\alpha |}%
R}(\theta _{1})\int\limits_{-\infty }^{\infty }\frac{d\theta _{2}}{2}%
e^{\theta _{2}-r\hat{m}e^{\theta _{2}}}\hat{W}_{\bar{\imath}}^{\mathbf{%
\alpha |}R}(\theta _{2})\right] ,  \label{corr}
\end{eqnarray}
where we have exploited the crossing relations stated in (\ref{ck}). Here
the functions $\hat{W}_{i}^{\mathbf{\alpha |}R}(\theta )$, $\hat{K}_{i}^{%
\mathbf{\alpha |}R}(\theta )$, $\ldots $ defined in (\ref{corr}) are the
massless limits of the corresponding functions $\hat{W}_{i}^{\mathbf{\alpha }%
}(\theta )$, $\hat{K}_{i}^{\mathbf{\alpha }}(\theta )$, $\ldots $ For all
the defects we will consider below, it turns out that the first contribution
to the previous correlation function is actually vanishing, so that (\ref
{corr}) is considerably simplified. In many of the examples we will treat
later, this is due to the fact that the amplitudes $\hat{K}_{i}^{\mathbf{%
\alpha }}(\theta )$ are vanishing in the first place, as a consequence of
the crossing relations (\ref{ck}). This will be the case for all energy
insensitive defects for which we will present a case-by-case computation of
the conductance below. The vanishing of the reflection part in (\ref{corr})
also occurs in some cases as a consequence of the parity of the function $%
\hat{K}_{i}^{\mathbf{\alpha }}(\theta )$. For instance, we find that, for
the energy operator defect such function, although initially non-vanishing,
satisfies $\hat{K}_{i}^{\mathbf{\alpha }}(\theta )=-\hat{K}_{i}^{\mathbf{%
\alpha }}(-\theta )$, such that $\lim_{m\rightarrow 0}\int_{-\infty
}^{\infty }d\theta \,\hat{K}_{i}^{\mathbf{\alpha }}(\theta )^{\ast }=0$.

\noindent We can now either use (\ref{corr}) in (\ref{kubod}) to compute the
conductance or evaluate the expression (\ref{conn}) directly in which the
frequency limit is already taken 
\begin{equation}
G^{\mathbf{\alpha }}(0)=\frac{1}{2(2\pi )^{3}}\sum\limits_{i}\int\limits_{-%
\infty }^{0}d\theta \,e^{\theta }\,w_{i\bar{\imath}}^{\mathbf{\alpha |}%
RR}[\ln (1-e^{\theta })_{,}\theta ]\,.
\end{equation}

There are, in addition, further generic results which can be obtained
independently of the specific defect. We present them at this stage and will
confirm their validity below by some specific examples. Specializing to the
case in which all $\ell $ defects are of the same type and equidistantly
separated, i.e., $y=y_{\alpha _{1}}=\cdots =y_{\alpha _{n}}$. As in the TBA
context (\ref{massl}), we can identify two distinct regions 
\begin{equation}
w_{i\bar{\imath}}^{\mathbf{\alpha |}RR}(\theta _{1,}\theta _{2})=\pi
^{2}\left\{ 
\begin{array}{l}
\overline{\hat{W}_{i}^{\mathbf{\alpha |}R}(\theta _{1})\hat{W}_{i}^{\mathbf{%
\alpha |}R}(\theta _{2})^{\ast }}\quad \quad \,\,\,\,\text{for finite }y \\ 
|\hat{W}_{i}^{\mathbf{\alpha |}R}|^{2}\qquad \qquad \qquad \qquad \,\text{%
for }y\rightarrow 0
\end{array}
\right.   \label{reg}
\end{equation}
where we used in addition (\ref{uss2}). Supported by our explicit examples
below, we find that for $y\rightarrow 0$ in (\ref{reg}) the amplitudes $%
\hat{W}_{i}^{\mathbf{\alpha |}R}(\theta )$ become independent functions of
the rapidity. As we have already argued above 
\begin{equation}
k_{ii}^{\mathbf{\alpha |}RR}(\theta _{1,}\theta _{2})=0.
\end{equation}
The two regions specified in (\ref{reg}) are in complete agreement with the
regions identified in equation (\ref{massl}), since we also consider here
the massless limit. When exploiting (\ref{reg}), our explicit examples below
yield the values of the conductance as those computed from (\ref{massl}). In
the regime $y\rightarrow 0$ this is very apparent, since when $|\hat{W}_{i}^{%
\mathbf{\alpha |}R}|=const$ it becomes equal to $2|T_{R}^{\mathbf{\alpha }}|$
and the conductance reduces just to the constant factor given by (\ref{reg})
times the value $1/2\pi $ obtained when defects are not present in the
theory. The vanishing of the first contribution in (\ref{corr}) is also
quite suggestive with regard to the TBA results, since the conductance
obtained in terms of thermodynamic quantities only involves the moduli of
the transmission or the reflection amplitudes, but not both simultaneously
and, in the light of the previous discussion, that seems to extend also to
the form factor computation. The coincidence in the regime for finite value
of $y$ between the Kubo formula based on (\ref{reg}) and the results from
the Landauer formula are less obvious and we support this by some explicit
computations for several specific defects, similar as in subsection 3.5.

\subsubsection{Energy insensitive defects, $\mathcal{D}^{0}(\bar{\protect\psi%
},\protect\psi )=0$, $\mathcal{D}^{\protect\beta }(\bar{\protect\psi},%
\protect\psi ),$ $\mathcal{D}^{\protect\gamma }(\bar{\protect\psi},\protect%
\psi )$, $\mathcal{D}^{\protect\delta _{\pm }}(\bar{\protect\psi},\protect%
\psi )$}

\vspace{-0.2cm}

\noindent A simple example to start with, which at the same time provides a
first test of the working of the Kubo formula in the UV-limit is a
transparent defect, i.e., purely transmitting. As shown in subsection 2.3.3,
examples for this are the absence of a defect $\mathcal{D}^{0}(\bar{\psi}%
,\psi )=0$ as well as the defect $\mathcal{D}^{\beta }(\bar{\psi},\psi )=g%
\bar{\psi}\gamma ^{1}\psi $, for which the associated reflection and
transmission amplitudes are given by (\ref{Rb}) and (\ref{Tb}). In this case
the observation $\hat{K}_{i}^{\mathbf{\alpha }}(\theta )=0$ is of course
trivial and therefore only the second integral in (\ref{corr}) is relevant.
The situation in which there is no defect was already commented on in the
paragraph after equation (\ref{:->}). We found in that case that the Kubo
formula leads to entirely consistent results with regard to our TBA
analysis, that is $G^{0}(0)=1/2\pi $ in the massless limit. Considering now
a theory with $n$ defects of the type $\mathcal{D}^{\beta _{i}}(\bar{\psi}%
,\psi )$, we find 
\begin{equation}
\hat{W}_{i}^{\beta _{1}\ldots \beta _{n}}(\theta )=2e^{-inB},\quad \quad 
\hat{K}_{i}^{\beta _{1}\ldots \beta _{n}}(\theta )=0\,,  \label{transp}
\end{equation}
simply by exploiting the expressions (\ref{Tb}) for a single defect and the
formulae (\ref{ttr}) and (\ref{ttr2}). From (\ref{transp}) it follows that 
\begin{equation}
w_{i\bar{\imath}}^{\beta _{1}\ldots \beta _{n}|RR}(\theta _{1},\theta
_{2})=4\pi ^{2},\qquad k_{ii}^{\beta _{1}\ldots \beta _{n}|RR}(\theta
_{1},\theta _{2})=0\,,  \label{grouse}
\end{equation}
just, as in the case in which the defect is absent. Therefore, we recover
once more the value 
\begin{equation}
G^{\beta _{1}\ldots \beta _{n}}(0)=\frac{1}{2\pi },  \label{nodef}
\end{equation}
for the conductance at zero mass and temperature.

The next complication arises for defects whose reflection and transmission
amplitudes are simultaneously non-vanishing, but at the same time are
independent functions of the rapidity. As we have seen in section 3.5, those
defects can be very easily handled in the context of a TBA calculation for
the conductance, since the modulus of the transmission amplitudes is a
constant, depending only on the defect coupling $g$. Therefore, the
conductance is simply given by the constant $|T^{\mathbf{\alpha }}|^{2}$
times the value (\ref{nodef}). The vanishing of the function $\hat{K}_{i}^{%
\mathbf{\alpha }}(\theta )$ can be established in those cases by exploiting
the crossing properties listened above. Namely, from (\ref{c1})-(\ref{c2})
we find $K_{i}=\tilde{K}_{i}^{\ast }=-\tilde{K}_{\bar{\imath}},$ whenever
the reflection amplitudes are independent of the rapidity, and therefore $%
\hat{K}_{i}^{\mathbf{\alpha }}=K_{i}+K_{i}^{\ast }=2\func{Re}(K_{i})$. The
latter quantity is zero for the defects $D^{\gamma }$ and $D^{\delta _{\pm
}} $ treated in section 2.3.4, when setting $\ y=0$, since the reflection
amplitudes are purely complex quantities. Therefore, the Kubo formula
computation leads to the same results as found in subsection 3.6.1, since we
find 
\begin{eqnarray}
w_{i\bar{\imath}}^{\gamma |RR}(\theta _{1},\theta _{2}) &=&4\pi ^{2}(1+4\tan
^{2}B/2),\,\,\,\,\,\,\,\,\,\,\,\,\,k_{ii}^{\gamma |RR}(\theta _{1},\theta
_{2})=0,\, \\
w_{i\bar{\imath}}^{\delta _{\pm }|RR}(\theta _{1},\theta _{2}) &=&\frac{4\pi
^{2}}{\cos ^{2}B},\,\,\,\,\,\qquad \qquad \qquad \,\,k_{ii}^{\delta _{\pm
}|RR}(\theta _{1},\theta _{2})=0\,,
\end{eqnarray}
which yields 
\begin{equation}
G^{\gamma }(0)=\frac{(1+4\tan ^{2}B/2)}{2\pi }\qquad \text{and\qquad }%
\,\,\,\,\,\,G^{\delta _{\pm }}(0)=\frac{1}{2\pi \cos ^{2}B}.
\end{equation}
As expected, for $B=0$ we recover once more the value (\ref{nodef}).

\subsubsection{The energy operator defect $\mathcal{D}(\bar{\protect\psi},%
\protect\psi )=g\bar{\protect\psi}\protect\psi $}

\vspace{-0.2cm}

\noindent Let us now treat the energy operator defect, the example which has
been most extensively studied in our previous sections. Considering first a
theory possessing a single defect of this type, we find 
\begin{equation}
\hat{W}_{i}^{\alpha }(\theta )=\frac{4\cos B\cosh ^{2}\theta }{\cosh 2\theta
+\cos 2B}\,\qquad \text{and\qquad }\hat{K}_{i}^{\alpha }(\theta )=\frac{%
2i\sin B\sinh \theta }{\sin B-\cosh \theta }.  \label{ggg}
\end{equation}
Therefore, in this case the amplitude $\hat{K}_{i}^{\mathbf{\alpha }}(\theta
)$ is non-vanishing. However, we find that $\hat{K}_{i}^{\mathbf{\alpha }%
}(\theta )=-\hat{K}_{i}^{\mathbf{\alpha }}(-\theta )$. This means that the
integral of this function (or its complex conjugated) is vanishing.
Consequently, only the transmission part contributes non-trivially to (\ref
{corr}). In order to evaluate (\ref{4p}) in the massless limit, we are
interested in this limit of (\ref{ggg}) which enters equation (\ref{corr}).
We obtain 
\begin{equation}
w_{i\bar{\imath}}^{\alpha |RR}(\theta _{1,}\theta _{2})=4\pi ^{2}\cos ^{2}B,
\label{gmass}
\end{equation}
which, together with (\ref{mf1}) leads to the result 
\begin{equation}
\left\langle J(r)Z_{\alpha }J(0)\right\rangle _{T=m=0}=\frac{\cos ^{2}B}{%
r^{2}}\,\,\Longrightarrow \,\,G^{\alpha }(0)=\frac{\cos ^{2}B}{2\pi },
\label{1d}
\end{equation}
again in agreement with the corresponding result (\ref{g1}) from the
Landauer formula.

Let us now proceed to the study of the conductance in the presence of a
double defect. Again, we consider first the case $T=m=0$ and two defects of
the energy operator type located at the origin and at a distance $y$ from
the origin, respectively. Expression (\ref{jzj}) again holds for that
situation with $n=2$. As explained in the paragraph after equation (\ref{tr2}%
), the Greek indices in the defect operator encode also the space
dependence. The reflection and transmission amplitudes are computed
according to (\ref{tr}) and (\ref{tr2}) with (\ref{ggg}). These functions
can thereafter be substituted into equation (\ref{4p}) in order to determine
the explicit form of the functions $w_{i\bar{\imath}}^{\alpha _{1}\alpha
_{2}|RR}$ and $k_{ii}^{\alpha _{1}\alpha _{2}|RR}$ in (\ref{4p}) for the
double defect system, which depend now on the distance $y$ between the
defects and their expressions become very cumbersome. Once more, it is
possible to show that the contribution to the conductance depending on $%
k_{ii}^{\alpha _{1}\alpha _{2}|RR}$ is vanishing and therefore only the
function $w_{i\bar{\imath}}^{\alpha _{1}\alpha _{2}|RR}$ will be of further
interest to us. However, it is relatively easy to show that in the massless
limit we find 
\begin{eqnarray}
w_{i\bar{\imath}}^{\alpha _{1}\alpha _{2}|RR}(\theta _{1,}\theta _{2})
&=&4\pi ^{2}\cos ^{4}B\left[ \frac{1+\cos (2\hat{m}ye^{\theta _{1}})\sin
^{2}B}{1+2\cos (2\hat{m}ye^{\theta _{1}})\sin ^{2}B+\sin ^{4}B}\right]  
\nonumber \\
&&\qquad \quad \times \left[ \frac{1+\cos (2\hat{m}ye^{\theta _{2}})\sin
^{2}B)}{1+2\cos (2\hat{m}ye^{\theta _{2}})\sin ^{2}B+\sin ^{4}B}\right] ,
\end{eqnarray}
such that we obtain 
\begin{eqnarray}
\left\langle J(r)Z_{\alpha _{1}}Z_{\alpha _{2}}J(0)\right\rangle _{T=m=0}\!
&=&\frac{\overline{w_{i\bar{\imath}}^{\alpha _{1}\alpha _{2}|RR}(\theta
_{1,}\theta _{2})}}{4\pi ^{2}r^{2}}=\frac{4\left[ 1+\sin ^{4}B\right] }{r^{2}%
\left[ \cos ^{2}(2B)-3\right] ^{2}},  \label{final1} \\
G^{\alpha _{1}\alpha _{2}}(0) &=&\frac{2}{\pi }\!\frac{1+\sin ^{4}B}{\left[
3-\cos ^{2}(2B)\right] ^{2}},  \label{final}
\end{eqnarray}
which precisely agrees with the corresponding result (\ref{o3}) obtained
from the Landauer formula. The overbar denotes as before the mean value of
the corresponding function.

As explained above, we can also predict the precise position of the second
plateau obtained within the TBA analysis given in equation (\ref{o1}). This
is achieved by considering \ previously to the UV-limit, the limit when the
distance between the defects $y\rightarrow 0$. By doing so we find 
\begin{equation}
\lim_{y\rightarrow 0}w_{i\bar{\imath}}^{\alpha _{1}\alpha _{2}|RR}(\theta
_{1,}\theta _{2})=\frac{4\pi ^{2}\cos ^{4}B}{(1+\sin ^{2}B)^{2}},
\label{4dy0}
\end{equation}
which gives 
\begin{eqnarray}
\,\,\lim_{y\rightarrow 0}\left\langle J(r)Z_{\alpha _{1}}Z_{\alpha
_{2}}J(0)\right\rangle _{T=m=0} &=&\frac{1}{r^{2}}\frac{\cos ^{4}B}{(1+\sin
^{2}B)^{2}}\,,\,  \label{y01} \\
\,\,\lim_{y\rightarrow 0}G^{\alpha _{1}\alpha _{2}}(0) &=&\frac{1}{2\pi }%
\frac{\cos ^{4}B}{(1+\sin ^{2}B)^{2}},  \label{y02}
\end{eqnarray}
in agreement with the value (\ref{o1}).

Finally, in order to match all the results in subsection 3.6, we would like
to address also the case $\ell =4$ in (\ref{kubod}), that is, we consider
now a complex free Fermion theory with four equidistant defects of the type $%
\mathcal{D}^{\alpha }(\bar{\psi},\psi )=g\bar{\psi}\psi $. As usual, we
denote their mutual distances by $y$. For the first region in (\ref{reg}),
that is the UV-limit, we find 
\begin{equation}
w_{i\bar{\imath}}^{\alpha _{1}\alpha _{2}\alpha _{3}\alpha _{4}|RR}(\theta
_{1,}\theta _{2})=\frac{f_{1}(\theta _{1})f_{1}(\theta _{2})}{(f_{2}(\theta
_{1})-f_{3}(\theta _{1}))(f_{2}(\theta _{2})-f_{3}(\theta _{2}))},
\label{mess}
\end{equation}
with 
\begin{eqnarray*}
f_{1}(\theta ) &=&(5-\cos 2B)\cos (4\hat{m}ye^{\theta })+2\cos (6\hat{m}%
ye^{\theta }))\sin ^{2}B)-128\pi ^{2}\cos ^{4}B(2+(6\cos (2\hat{m}ye^{\theta
}) \\
f_{2}(\theta ) &=&1192\cos 2B-348\cos 4B+24\cos 6B-\cos 8B-995-256\sin
^{2}B\cos (6\hat{m}ye^{\theta })) \\
f_{3}(\theta ) &=&128\sin ^{2}B((17-12\cos 2B+\cos 4B)\cos (2\hat{m}%
ye^{\theta })-4(\cos 2B-2)\cos (4\hat{m}ye^{\theta })\,.
\end{eqnarray*}

\noindent This expression appears somewhat messy, but when proceeding as
indicated in (\ref{reg}) it will simplify considerably. Computing the mean
value of this function we find

\begin{eqnarray}
\left\langle J(r)Z_{\alpha _{1}}Z_{\alpha _{2}}Z_{\alpha _{3}}Z_{\alpha
_{4}}J(0)\right\rangle _{T=m=0}\! &=&\frac{\overline{w_{RR}^{\alpha
_{1}\alpha _{2}\alpha _{3}\alpha _{4}}(\theta _{1,}\theta _{2})}}{4\pi
^{2}r^{2}}  \nonumber \\
&=&\frac{1}{2r^{2}}\left[ 1+\frac{\cos ^{8}B}{[\cos ^{4}B-2(1+\sin
^{2}B)^{2}]^{2}}\right] , \\
G^{\alpha _{1}\alpha _{2}\alpha _{3}\alpha _{4}}(0) &=&\frac{1}{4\pi }\left(
1+\frac{\cos ^{8}B}{[\cos ^{4}B-2(1+\sin ^{2}B)^{2}]^{2}}\right) ,
\end{eqnarray}
in complete agreement with the corresponding TBA value (\ref{o4}). We can
also predict the precise position of the second plateau which, according to (%
\ref{reg}) is expected for the conductance. Once more we find complete
agreement with the outcome of our TBA analysis, since in this case 
\begin{equation}
\lim_{y\rightarrow 0}w_{i\bar{\imath}}^{\alpha _{1}\alpha _{2}\alpha
_{3}\alpha _{4}}(\theta _{1,}\theta _{2})=\left( \frac{2\pi \cos ^{4}B}{\cos
^{4}B-2(1+\sin ^{2}B)^{2}}\right) ^{2}\,,
\end{equation}
which gives 
\begin{eqnarray}
\,\lim_{y\rightarrow 0}\left\langle J(r)Z_{\alpha _{1}}Z_{\alpha
_{2}}Z_{\alpha _{3}}Z_{\alpha _{4}}J(0)\right\rangle _{T=m=0} &=&\frac{1}{%
r^{2}}\left( \frac{\cos ^{4}B}{\cos ^{4}B-2(1+\sin ^{2}B)^{2}}\right)
^{2}\,\,,\,  \label{cor2} \\
\,\,\lim_{y\rightarrow 0}G^{\alpha _{1}\alpha _{2}\alpha _{3}\alpha _{4}}(0)
&=&\frac{1}{2\pi }\left( \frac{\cos ^{4}B}{\cos ^{4}B-2(1+\sin ^{2}B)^{2}}%
\right) ^{2}\,,  \label{con2}
\end{eqnarray}
that is, the same expression as (\ref{o2}).

\section{Conclusions}

\vspace{-0.2cm} \setcounter{equation}{0}

We have exploited the special features of 1+1 dimensional integrable quantum
field theories in order to compute the DC\ conductance in an impurity
system. For this purpose several non-perturbative techniques have been used.
As the main tools we employed the thermodynamic Bethe ansatz in a Landauer
transport theory computation and the form factor expansion in the Kubo
formula. 

\emph{The comparison between the Landauer formula (\ref{1}) and the Kubo
formula (\ref{2}) yields in particular an identical plateau structure for
the DC conductance in the ultraviolet limit. }

We have explained to what extend integrability can be exploited in order to
determine the reflection and transmission amplitudes through a defect.
Unfortunately, for the most interesting situation in this context, namely
when $R/\tilde{R}$ and $T/\tilde{T}$ are simultaneously non-vanishing, the
Yang-Baxter-bootstrap equations narrow down the possible bulk theories to
those which possess rapidity independent scattering matrices \cite{DMS,CFG}.
By means of a relativistic potential scattering theory we compute for
several types of defects the $R/\tilde{R}$s and $T/\tilde{T}$s, thus
enlarging the set of examples available at present. We confirm that for real
potentials parity is preserved, but otherwise essentially all possible
combinations of parity breaking can occur. From the knowledge of the single
defect amplitudes the multiple defect amplitudes, which exhibit the most
interesting physical behaviours, can be computed in a standard fashion \cite
{CT,Merz}.

We newly formulate the TBA equations for a defect with simultaneously
non-vanishing reflection and transmission amplitudes. We indicate how these
equations can be used to compute various thermodynamic quantities, which
are, however, most interesting only when considered per unit length. By
means of the TBA we compute the density distribution functions and use them
to evaluate the Landauer conductance formula (\ref{1}) for various defects
in a complex free fermionic theory. We predict analytically the most
prominent features in the conductance as a function of the temperature, i.e.
the plateaux.

We formulate the Kubo formula \cite{kubo} for a situation in which defects
are present (\ref{2}). We evaluate the current-current correlation functions
occurring in there by means of another non-perturbative method based on
integrability, namely the bootstrap form factor approach \cite{KW,Smir}. We
provide closed formulae which solve explicitly the defect recursive
equations involving any arbitrary number of particles. As for the Landauer
formula, we also predict in this case the plateaux in the conductance as a
function of the temperature analytically.

There are several interesting open issues. Most challenging is to treat in
full generality the massive and temperature dependent case of (\ref{2}).
Unfortunately, the formulation of non-perturbative methods do not yet cover
that situation \cite{CF11} and it remains to be clarified how the form
factor bootstrap program for the computation of two-point functions can be
extended to that case. It would be further interesting to compute
thermodynamic quantities per unit length by means of the TBA formulated in
section 3.3. To classify possible defects more systematically is desirable
even for free theories.\medskip 

\noindent \textbf{Acknowledgments: }We are grateful to the Deutsche
Forschungsgemeinschaft (Sfb288), for financial support. We thank F.
G\"{o}hmann for discussions.

\end{document}